\documentclass[prd,preprintnumbers,nofootinbib]{revtex4} 

\usepackage{color}
\usepackage{epsfig}
\usepackage{graphicx}
\usepackage{amsmath}
\usepackage{amssymb}
\usepackage{slashed}
\newcommand{\beq}{\begin{equation}}
\newcommand{\eeq}{\end{equation}}
\newcommand{\bea}{\begin{eqnarray}}
\newcommand{\eea}{\end{eqnarray}}
\newcommand{\bem}{\begin{multline}}
\newcommand{\eem}{\end{multline}}
\newcommand{\beg}{\begin{gather}}
\newcommand{\eeg}{\end{gather}}

\def\eq#1{{Eq.~(\ref{#1})}}

\newcommand{\ben}{\begin{eqnarray*}}
\newcommand{\een}{\end{eqnarray*}}

\def\gsim{ \,\, \vcenter{\hbox{$\buildrel{\displaystyle >}\over\sim$}}
 \,\,}
\def\lsim{ \,\, \vcenter{\hbox{$\buildrel{\displaystyle <}\over\sim$}}
 \,\,}

\begin{document}
\title{{\bf CGC predictions for p+Pb collisions at the LHC \\[0cm] }}

\preprint{RBRC 969}

\author{ {\bf Javier L. Albacete$^a$\thanks{e-mail:
      albacete@ipno.in2p3.fr},\hspace{0.1cm} Adrian
    Dumitru$^b$\thanks{Adrian.Dumitru@baruch.cuny.edu}, Hirotsugu
    Fujii$^c$\thanks{hfujii@phys.c.u-tokyo.ac.jp} and Yasushi
    Nara$^d$\thanks{nara@aiu.ac.jp}} \\[0.15cm] 
{\it \small
    $^a$IPNO, Universit\'e Paris-Sud 11, CNRS/IN2P3, 91406 Orsay, France}\\ {\it \small $^b$Department of
    Natural Sciences, Baruch College, CUNY, 17 Lexington Avenue, New
    York, NY 10010, USA}\\ {\it \small RIKEN BNL Research Center,
    Brookhaven National Laboratory, Upton, NY 11973, USA }\\ {\it
    \small $^{c}$ Institute of Physics, University of Tokyo, Komaba,
    Tokyo 153-8902, Japan}\\ {\it \small $^{d}$ Akita International
    University, Yuwa, Akita-city 010-1292, Japan} }

\begin{abstract}

We present predictions for multiplicities and single inclusive
particle production in proton-lead collisions at the LHC. The main
dynamical input in our calculations is the use of solutions of the
running coupling Balitsky-Kovchegov equation tested in e+p data. These
are incorporated into a realistic model for the nuclear geometry
including fluctuations of the nucleon configurations. Particle production
is computed via either $k_t$-factorization or the {\it hybrid}
formalisms to obtain spectra and yields in the central and forward
rapidity regions, respectively. These baseline predictions will be
useful for testing our current understanding of the dynamics of very
strong color fields against upcoming LHC data.
\end{abstract}

\maketitle

\section{Introduction}

Heavy-ion collision programs performed previously at the RHIC and SPS
accelerators have always benefited tremendously from insight obtained
by proton (deuteron)-nucleus collisions. It is expected, also, that
the planned p+Pb run at the LHC will provide crucial benchmarks for
the characterization of the hot QCD medium produced in lead-lead
collisions, arguably a Quark Gluon Plasma (QGP).  The absence of
intricate final state effects induced by the formation of a QGP
renders proton-nucleus collisions an excellent laboratory to study the
so called initial state effects. These originate from the fact that
the colliding nuclei do not behave as a mere incoherent superposition
of their constituent nucleons. Rather, coherence effects are
important and modify not only the partonic flux into the collision,
but also the underlying dynamics of particle production in the
scattering processes. A careful distinction of initial from final
state effects is of vital importance for a proper characterization of
the latter, as they may lead sometimes to qualitatively similar
phenomena in observables of interest. Physics prospects for the p+Pb
at the LHC run were recently compiled in \cite{Salgado:2011wc}.

Besides its role as a reference experiment, the p+Pb run at the LHC
will provide access to kinematic regions never explored so far in
nuclear collisions and thus carries great potential for discovery of
new QCD phenomena on its own. In particular, the huge leap forward in
collision energy with respect to previous high energy electron-nucleus
or proton-nucleus experiments\footnote{The expected center of mass
  collision energy for the p+Pb run is 5~TeV, to be compared to a
  maximal energy of 200 GeV for RHIC d+Au collisions.} will probe the
nuclear wave function at values of Bjorken-$x$ smaller than ever
before. It is theoretically well established that at small enough
values of Bjorken-$x$ QCD enters a novel regime governed by large
gluon densities and non-linear coherence
phenomena~\cite{Mueller:1999wm}.  The Color Glass Condensate (CGC)
effective theory provides a consistent framework to study QCD
scattering at small-$x$ or high collision energies (for a review see
e.g.  \cite{Gelis:2010nm,Weigert:2005us}). It is based on three main
physical ingredients: First, high gluon densities correspond to strong
classical fields~\cite{McLerran:1993ni,McLerran:1993ka}, which permit
ab-initio first principles calculation of ``wave functions'' at small
$x$ through classical techniques. Next, quantum corrections are
incorporated via non-linear renormalization group equations such as
the B-JIMWLK hierarchy or, in the large-$N_{c}$ limit, the BK
equation~\cite{Balitsky:1996ub,Kovchegov:1999yj} which describe the
evolution of the hadron wave function towards small $x$. The
non-linear, density-dependent terms in the CGC evolution equations are
ultimately related to unitarity of the theory and, in the appropriate
frame and gauge, can be interpreted as due to gluon recombination
processes that tame or {\it saturate} the growth of gluon densities
for modes with transverse momenta below a dynamically generated scale
known as the saturation scale, $Q_{s}(x)$. Finally, the presence of
strong color fields $\mathcal{A}\sim1/g$ leads to breakdown of
standard perturbative techniques to describe particle production
processes based on a series expansion in powers of the strong coupling
$g$. Terms of order $g\mathcal{A} \sim \mathcal{O}(1)$ need to be
resummed to all orders. The CGC provides the tools to perform such
resummation although the precise prescription for the resummation may
vary from process to process or colliding system.

An important step towards promoting the CGC to a practical
phenomenological tool has been performed recently through the
calculation of higher order corrections to the BK-JIMWKL
equations~\cite{Kovchegov:2006vj,Balitsky:2006wa, Gardi:2006rp,
  Balitsky:2008zz}. In particular, the calculation of running coupling
corrections to the evolution kernel of the BK equation (referred to as
rcBK henceforth) made it possible to describe various data at high
energies in terms of solutions of the rcBK equation, thus closing the
gap between first principles theory calculations and data
\cite{Albacete:2007yr}. 

Despite the fact that CGC effects are expected to be enhanced in
nuclei versus protons, which is due to the larger valence charge
densities per unit transverse area in nuclei, so far the most
exhaustive searches of the saturation phenomenon have been performed
using data on proton reactions. This is mainly due to the large body
of high quality DIS data on protons at small-$x$. Thus, the global
fits performed by the AAMQS collaboration in \cite{Albacete:2009fh,
  Albacete:2010sy, Albacete:2012rx} demostrated that the rcBK equation
successfully accounts for the $x$-dependence of inclusive structure
functions measured in electron-proton collisions at HERA at small-$x$
(see also \cite{Kuokkanen:2011je} for fits to diffractive data). Here,
we shall use the resulting AAMQS parametrizations of the dipole-proton
scattering amplitude as a basic building block for the nuclear
unintegrated distributions. Furthermore, as discussed in more detail
below, single inclusive $p_\perp$ spectra in proton+proton collisions
at Tevatron, LHC and forward RHIC data are well described in the CGC
framework~\cite{Albacete:2010bs, Fujii:2011fh, Fujii:2012zza}. In case
of heavy-ion collisions, the most compelling indications of CGC
effects in data are due to suppression phenomena observed at RHIC d+Au
collisions in the forward rapidity region. Actually, single particle
distributions in both p+p and d+Au collisions and the depletion of
nuclear modification factors $R_{dAu}$ were well described by CGC
calculations in \cite{Albacete:2010bs, Fujii:2011fh}. However, forward
RHIC data are close to the kinematic limit of phase space, probing the
projectile wave function at large values of $x\to1$ and the
fragmentation functions at $z\to1$. This allows for alternative
descriptions of the data based on large-$x$ phenomena such as energy
loss~\cite{Kopeliovich:2005ym}. A clearer signal of gluon saturation
is provided by the observed disappearance of back-to-back di-hadron
correlations in d+Au collisions. CGC calculations predicted first
\cite{Marquet:2007vb}, and later on quantitatively accounted
for~\cite{Albacete:2010pg, Stasto:2011ru}, such suppression. The
disappearance of the away-side peak can be understood in terms of
multiple scatterings controlled by a dynamical transverse scale, the
saturation scale of the nucleus. Even though no description of such
data that does not invoke saturation effects is known to date (see
however \cite{Kang:2011bp}), it has been argued that large-$x$ effects
like energy loss or enhancement of double scattering processes may
blur the interpretation of angular decorrelation between hadron pairs
as due to CGC effects~\cite{Strikman:2010bg}.  Additionally, the good
description of the energy and centrality dependence of integrated
multiplicities in RHIC Au+Au and d+Au collisions and in Pb+Pb
collisions at the LHC in CGC models lends further support to the idea
that saturation effects are relevant in present data, although the
larger degree of modeling involved in their calculation prevents
making any definite claims. In summary, although it may be argued that
the present data do not provide sharp and direct evidence for neither
the saturation regime of QCD nor the CGC approach to saturation, it is
fair to say that HERA, RHIC and available LHC data lead to a rather
coherent picture, with several phenomena finding their natural
interpretation in terms of high density gluonic systems together with
a consistent quantitative description in the CGC framework at its
present degree of accuracy.

Regardless of the question whether the CGC is the most suited
framework for their description there is broad consensus that
coherence effects are important for the interpretation of present data
on heavy ion collisions. In fact most --if not all-- of the different
phenomenological approaches for the description of particle production
--both in the soft or hard sector-- include physical ingredients
related to either the depletion of gluon content of nuclei (e.g.\ the
leading twist shadowing of the nuclear parton distributions in the
collinear factorization formalism or string fusion processes in Dual
Parton or percolation models), the breakdown of independent particle
production from the individual participant nucleons (e.g.\ through the
presence of energy dependent cut-offs in event generators like HIJING
or HYDJET) or effective resummations of multiple
scatterings~\footnote{Here we just mention a few well known
  examples; a rather exhaustive compilation of phenomenological works
   for the description of particle production in HIC can be found in
  e.g \cite{Abreu:2007kv}.}. All these ingredients are akin, at least at a
conceptual level, to those dynamically built in the CGC, although they
are formulated in very different ways.

The p+Pb run at the LHC will provide an excellent --and probably in
the near future unique-- possibility to disentangle the presently
inconclusive situation on the role of CGC effects and also to
distinguish among different approaches to describe high energy
scattering in nuclear collisions. On the one hand, the LHC shall bring
us closer to the limit of asymptotically high energy in which the CGC
formalism is developed, thus reducing theoretical uncertainties on its
applicability. Equivalently, the value of the saturation scale is
expected to be a factor $\sim2\div4$ times larger than at RHIC, so
saturation effects should be visible in a larger range of transverse
momenta, deeper into the perturbative domain. On the other hand, the
much extended energy reach of the LHC will allow measurements far
from the kinematic limit up to very forward rapidities, thus
minimizing the role of large-$x$ effects which obscured the
interpretation of forward RHIC data.

Although more exclusive observables like di-hadron or hadron-photon
correlations are expected to better discriminate between different
approaches, a first test for models of particle production
shall come from data on inclusive multiplicities and single particle
distributions as they are much easier to obtain experimentally. In
this work we aim to provide up-to-date predictions for these
observables within the framework of the running coupling BK Monte Carlo
(rcBK-MC) previously presented in \cite{Albacete:2010ad} and built as
an upgrade of the original KLN-MC code \cite{Drescher:2007ax}. The
rcBK-MC model presented here attempts to incorporate
up-to-date tools on the CGC theoretical side, namely the use of
solutions of the rcBK equation as main dynamical input to describe the
$x$ and transverse momentum dependence of nuclear unintegrated gluon
distributions, combined with the empiric information gained from the
analysis of e+p, p+p and RHIC nuclear collisions. The Monte Carlo
treatment allows for a realistic treatment of the collision geometry
including in particular realistic nuclear density distributions as
well as fluctuations of the configurations of nucleons. This is
important in order to obtain the number of collisions $N_{\rm coll}$
used in the definition of the nuclear modification factor $R_{pA}$
consistently with the computed p+A spectra, within one single
framework. Also, fluctuations of the number of target nucleons are
particularly relevant for particle production in the high
density regime where the number of small-$x$ gluons is not linearly
proportional to the number of valence charges.

There are two distinct but related approaches to hadron production in
high energy asymmetric (such as proton-nucleus or very forward
proton-proton and nucleus-nucleus) collisions.  In such collisions,
particle production processes in the central rapidity region probe the
wave functions of both projectile and target at small values of
$x$. Here, one may employ the $k_t$-factorization formalism where both
the projectile and target are characterized in terms of their rcBK
evolved unintegrated gluon distributions (UGDs). 

At more forward rapidities, on the other hand, the proton is probed at
larger values of $x$ while the target nucleus is shifted deeper into
the small-$x$ regime. Here, $k_{t}$-factorization fails to grasp the
dominant contribution to the scattering process. Rather, the {\it
  hybrid} formalism proposed in ref.~\cite{Dumitru:2005gt} and
embedded recently in a Monte Carlo treatment of the nuclear geometry
\cite{Fujii:2011fh,Fujii:2012zza} is more appropriate. In the hybrid
formalism the large-$x$ degrees of freedom of the proton are described
in terms of usual parton distribution functions (PDFs) of collinear
factorization which satisfy the momentum sum rule exactly and which
exhibit a scale dependence given by the DGLAP evolution equations. On
the other hand, the small-$x$ glue of the nucleus is still described
in terms of its UGD. Recently the hybrid formalism has been improved
through the calculation of inelastic contributions that may become
important at high transverse momentum~\cite{Altinoluk:2011qy}. We
shall investigate the effect of this new production channel in our
predictions.  Unfortunately, no smooth matching between the
$k_{t}$-factorization and {\it hybrid} formalisms is known to date.
Also, their corresponding limits of applicability --equivalently the
precise value of $x$ at which one should switch from one to the
other-- have only been estimated on an empirical basis. We shall
explore the stability of CGC predictions using one or the other
formalism as one varies the kinematics of the detected hadron.

One important goal of this work is to systematically assess the model (and
implementation) uncertainties. Such calibration is mandatory in order to
actually prove or disprove different approaches by comparison to
data. This is specially so in the case of semi-inclusive observables
such as the ones discussed in this work where ``only''
quantitative differences arise between predictions from different
approaches (contrary to the case of more exclusive observables).  

In the rcBK-MC model presented here, uncertainties originate from
either the modeling of non-perturbative aspects of the collision or
from high-order corrections to fixed order perturbative calculations
for particle production. Among the former are the initial conditions
for rcBK evolution, which are only partially constrained by global
fits to e+p data, or modeling of the impact parameter dependence of
the nuclear UGDs. Perturbative tools such as the rcBK equation are
ill-suited to provide a simultaneous, unified description of both the
Bjorken-$x$ and impact parameter dependence of the UGD of a nucleus
(or any other hadron) since the latter is related to the physics of
confinement. This makes some degree of modeling
unavoidable. Concerning the uncertainties due to fixed order
perturbative calculations, they are either absorbed in $K$-factors or
explored through the variation of factorization and running coupling
scales. The $K$-factors discussed here should also absorb effects not
included in the CGC approach.

\section{The nuclear unintegrated gluon distributions}

The paucity of small-$x$ data with nuclei prevents a direct
empiric determination of the nuclear unintegrated gluon distribution
at small-$x$. Here we shall build the nuclear UGD entirely from the
nucleon UGD extracted by the AAMQS collaboration through global fits
to inclusive e+p data on structure functions.

Before proceeding further, let us briefly recall the basic features of
the AAMQS approach. The main dynamical ingredient is the dipole
formulation of deep inelastic scattering (DIS). The AAMQS fits rely
upon is the $q\bar{q}$ dipole scattering amplitude off a hadron,
$\mathcal{N}_{F}(r,x, {\bf R})$, where $x$ is the usual Bjorken
scaling variable in DIS, $r$ is the dipole transverse size and ${\bf
  R}$ the transverse position at which the hadron target is
probed. The index $F$ refers to the fact that the dipole constituents
--quark and antiquark-- belong to the fundamental representation of
$SU(3)$. In the large $N_c$-limit the B-JIMWLK equations reduce to a
single, closed equation for the $x$-dependence of the dipole
amplitude, which is the so-called BK equation.  Under the assumption
of translational invariance implicit in the AAMQS approach, and
hence omitting the ${\bf R}$-dependence of the dipole amplitude
accordingly, the BK equation equation including running coupling
corrections (referred to as rcBK in what follows) reads
\begin{equation}
  \frac{\partial\mathcal{N}_{F}(r,x)}{\partial\ln(x_0/x)}=\int d^2{\underline r_1}\
  K^{run}({\underline r},{\underline r_1},{\underline r_2})
  \left[\mathcal{N}_{F}(r_1,x)+\mathcal{N}_{F}(r_2,x)
-\mathcal{N}_{F}(r,x)-\mathcal{N}_{F}(r_1,x)\,\mathcal{N}_{F}(r_2,x)\right]\,
\label{bk1}
\end{equation}
where ${\bf r=r_{1}+r_{2}}$ \footnote{We use the notation $v\equiv |\bf v|$ for two-dimensional vectors throughout the paper.} and $\mathcal{K}^{run}$ is the evolution kernel including running coupling corrections:
\begin{equation}
  K^{{run}}({\bf r},{\bf r_1},{\bf r_2})=\frac{N_c\,\alpha_s(r^2)}{2\pi^2}
  \left[\frac{1}{r_1^2}\left(\frac{\alpha_s(r_1^2)}{\alpha_s(r_2^2)}-1\right)+
    \frac{r^2}{r_1^2\,r_2^2}+\frac{1}{r_2^2}\left(\frac{\alpha_s(r_2^2)}{\alpha_s(r_1^2)}-1\right) \right]\,.
\label{kbal}
\end{equation}
In practical implementations, the running coupling in Eq.~(\ref{kbal})
is regularized in the infrared by freezing it to a constant value
$\alpha_{\rm fr}=0.7$.

Solving the BK equation is an initial value problem, i.e.\ it is well
defined only after initial conditions at the initial evolution scale,
$x_{0}=10^{-2}$ in the AAMQS fits, and for all values of the
dipole size $r$ have been provided. This introduces free parameters,
ultimately of non-perturbative origin, to be fitted to data. In the
AAMQS rcBK fits to HERA data the initial conditions are taken in the
form
\begin{equation}
\mathcal{N}_{F}(r,x\!=\!x_0)=
1-\exp\left[-\frac{\left(r^2\,Q_{s0,{\rm proton}}^2\right)^{\gamma}}{4}\,
  \ln\left(\frac{1}{\Lambda\,r}+e\right)\right]\ ,
\label{ic}
\end{equation}
where $\Lambda=0.241$ GeV, $Q^{2}_{s0,{\rm proton}}$ is the saturation
scale at the initial scale $x_{0}$ and $\gamma$ is a dimensionless
parameter that controls the steepness of the tail of the unintegrated
gluon distribution for momenta above the saturation scale
$k_t>Q_{s0,{\rm proton}}$. Both $Q_{s0,{\rm proton}}^{2}$ and $\gamma$
are fitted to data. Although the AAMQS fits clearly favor values
$\gamma>1$, they do not uniquely determine its optimal value (and
neither does the analysis of forward RHIC data performed in
ref.~\cite{Fujii:2011fh}). Rather, different pairs of $(Q_{s0,{\rm
    proton}}^{2},\gamma)$ parameters provide comparable values of
$\chi^{2}/{\rm d.o.f}\sim1$. The reason for this behavior is that they
are correlated with other parameters, such as the overall
normalization of the $\gamma^*$-p cross section, and also that HERA
data is too inclusive to constrain exclusive features of the proton
UGD.  In order to account for such uncertainty we shall consider two
of the AAMQS sets, corresponding to $(Q_{s0,{\rm
    proton}}^2,\gamma)$=(0.168 GeV$^2$, 1.119) and (0.157 GeV$^2$,
1.101). Additionally, we shall also consider the McLerran Venugopalan
(MV) model~\cite{McLerran:1993ni,McLerran:1993ka} which corresponds to
Eq.~(\ref{ic}) with $\gamma=1$; the MV model is well established
theoretically (for very large nuclei, $gA^{1/3}\gg1$) and has been
used frequently in the literature. We note that $\gamma>1$ for the
proton may arise from corrections to the effective action of the MV
model of higher order in the valence color charge
density~\cite{Dumitru:2011ax}. Such corrections are expected to
decrease with increasing nuclear thickness. Therefore, it is
conceivable that the dipole-nucleus scattering amplitude may be better
represented by the MV model than by initial conditions with
$\gamma>1$. However, using the AAMQS ($\gamma>1$) i.c.\ for the proton
simultaneously with an MV i.c.\ for the nucleus leads to a monotonical
increase (with $p_{t}$) of the nuclear modification factor $R_{\rm
  pPb}$ which is due to the different power-law fall off of the
respective UGDs. We shall not study this option in detail in this
work.
Also, we recall that in the AAMQS fits the strong coupling is evaluated according to the following expression:
\begin{equation}\label{eq:alpha}
\alpha_{s}(r^{2})=\frac{4\pi}{\beta
\ln\left(\frac{4C^{2}}{r^{2}\Lambda^{2}}+\mu\right)}
\end{equation}
where $\beta=11-\frac{2}{3}N_{f}$, $N_{f}=3$, $\Lambda=0.241$ GeV
and the constant $C$ under the logarithm accounts for the uncertainty
inherent to the Fourier transform from momentum space,
where the original calculation was performed. 
A parameter $\mu$ is introduced to regulate the strong coupling
for large dipole sizes, and fixed by the condition $\alpha_s(\infty)=\alpha_{fr}$.
 The $(Q_{s0,{\rm proton}}^{2},\gamma, \alpha_{fr}, C)$-values considered here
are shown in Table~\ref{tabfits}. 
%

\begin{table}[htbp]
   \centering
   \begin{tabular}{|c|c|c|c|c|c|c|c|c|}\hline
UGD Set  & $ Q_{s0,{\rm proton}}^{2}$ (GeV$^{2}$)  & $\gamma$ & $\alpha_{fr}$ & $C$   \\\hline
 MV    & 0.2 & 1 &  0.5 & 1\\
 g1.119     & 0.168 & 1.119  & 1.0 & 2.47  \\
 g1.101   & 0.157  & 1.101  & 0.8  & 1 \\ \hline
   \end{tabular}
   \caption{\small Summary of parameters for the three
   dipole-proton scattering amplitudes (or UGDs) considered in this work. }
   \label{tabfits}
\end{table}

In the large-$N_{c}$ limit the gluon dipole scattering amplitude
required for the unintegrated gluon distributions can be obtained from
the quark dipole scattering amplitude that solves the rcBK equation:
\begin{equation}
\mathcal{N}_A(r,x,{\bf R})=2\,\mathcal{N}_{F}(r,x,{\bf
  R})-\mathcal{N}_{F}^2(r,x,{\bf R})\,
\label{NA}
\end{equation}
where the subscript $A$ simply refers to the fact that gluons belong
to the adjoint representation of $SU(3)$. Note that this relation
entails that the saturation momentum relevant for gluon scattering is
larger than that for quark scattering by about a factor of 2, at the
initial rapidity $x=x_0$.

The nuclear UGDs needed for particle production in the
$k_{t}$-factorization and {\it hybrid} frameworks are related to the
quark and gluon dipole scattering amplitudes via 2-dimensional Fourier
transforms. In particular, the UGD entering the $k_t$-factorization
formula is given by
\begin{equation}
\varphi(k,x,{\bf R})=\frac{C_F}{\alpha_s(k)\,(2\pi)^3}\int d^2{\bf r}\
e^{-i{\bf k}\cdot{\bf r}}\,\nabla^2_{\bf r}\,\mathcal{N}_A(r,x,{\bf R})\,.
\label{phi}
\end{equation}
The function $\varphi$ is dimensionless and corresponds to the number
of gluons per unit transverse area and per transverse momentum space cell.

Eq.~(\ref{phi}) was written originally for fixed coupling. In order to
be consistent with our treatment of the small-$x$ evolution, we have
extended it by allowing the coupling in the denominator to run with
the momentum scale. As explained in \cite{Albacete:2010ad}, this
modification turns out to be important to obtain a good description of
the centrality dependence of the charged hadron multiplicities
measured in Au+Au and Pb+Pb collisions at RHIC and the LHC
respectively, which are otherwise too flat. In turn, the
following UGDs in the fundamental and adjoint representations are
needed in the {\it hybrid} framework:
\begin{equation}
\widetilde{\cal N}_{F(A)}(k,x,{\bf R})=\int d^2{\bf r}\,e^{-i{\bf k}\cdot{\bf
    r}}\left[1-\mathcal{N}_{F(A)}(r,x,{\bf R})\right]~.
\label{phihyb}
\end{equation}
Note that although the AAMQS approach assumes translational invariance
--i.e.\ impact parameter independence-- of the dipole scattering
amplitude over transverse distances of the order of the nucleon
radius, $R_{N}$, we have made explicit such dependence in equations
Eqs.~(\ref{NA}-\ref{phihyb}) for consistency with the notation
employed for nuclear UGDs where the impact parameter dependence must be
considered.

Let us now discuss our model for the nuclear UGD, starting
from the one for a nucleon\footnote{Strictly speaking, the AAMQS fits
  provide information only on the proton UGD or dipole scattering
  amplitude. We shall assume that isospin effects are negligible for
  the gluon distributions at small-$x$ and consider it equal to the
  one of a neutron.}. First, we assume that the functional form of the
quark dipole scattering amplitude off a nucleus at the initial
saturation scale $x=x_{0}=0.01$ is the same as for the quark dipole
scattering amplitude off a proton but with a shifted initial
saturation scale that depends on the local density at every point in
the transverse plane, ${\bf R}$. In other words, we shall
replace
\begin{equation}
Q_{s0,{\rm proton}}^{2}\to Q_{s0,{\rm nucleus}}^{2}({\bf R})
\label{qsnucl}
\end{equation}
in Eq. (\ref{ic}) in order to define the initial conditions for the
evolution of the quark dipole scattering off a nucleus at every
transverse point in the nucleus. Then, the initial conditions given
by Eqs.~(\ref{ic}) and (\ref{qsnucl}) are evolved locally using the
impact parameter independent rcBK evolution defined by
Eqs.~(\ref{bk1}-\ref{kbal}). This provides the full $(r,x)$-dependence
of the quark dipole-nucleus scattering amplitude at every point in the
transverse plane of the nucleus. Finally,
Eqs.~(\ref{phi}-\ref{phihyb}) are used to calculate their Fourier
transform which provide the complete transverse momentum $k_t$,
Bjorken-$x$ and impact parameter (${\bf R}$) dependence of the nuclear
UGDs.

To complete our discussion of the initial conditions we explain how we
construct $Q_{s0,{\rm nucleus}}^{2}({\bf R})$. We treat the transverse
positions of nucleons as random variables following a two-dimensional
projection of the Woods-Saxon distribution, $T_{A}({\bf R})$.  Each
configuration consist of a list of random coordinates ${\bf r}_i$,
$i=1\dots A$, for the locations of the different nucleons in the
transverse plane; $A$ denotes the atomic mass number of the
nucleus. Multi-nucleon correlations are neglected except for imposing
a short-distance hard core repulsion which enforces a minimal distance
$\approx 0.4$~fm between any two nucleons.  

Every such configuration defines a different local density in the
transverse plane of the nucleus. Obviously, the smallest non-zero
local density corresponds to the presence of a single nucleon. On the
other hand, rare fluctuations where a large number of nucleons is
encountered at a given transverse position can occur. Such configurations
correspond to a high initial saturation scale, $Q^{2}_{s0,{\rm
    nucleus}}({\bf R})$.  For a given configuration, the initial
saturation momentum $Q^{2}_{s0,{\rm nucleus}}({\bf R})$ at the
transverse coordinate ${\bf R}$ is taken to be proportional to the
local variance of the density of valence charges which is itself
proportional to the number of overlapping nucleons at that transverse
point, $N({\bf R})$:
\begin{equation}  \label{eq:addQs2ini}
a)\quad Q_{s0,{\rm nucleus}}^2({\bf R}) = N({\bf R}) \, Q_{s0,\, {\rm proton}}^2~.
\end{equation}
We shall refer to this option as {\it natural} in what follows.
Note that these valence charge sources generate the small-$x$ gluon
fields described by the UGD {\em coherently}. For the AAMQS initial
condition, the prescription~(\ref{eq:addQs2ini}) may lead to an
inconsistent definition of the nuclear modification factor in the
limit of large transverse momentum because additivity of the dipole
scattering amplitude in the number of nucleons, ${\cal N}(r;x_0,{\bf
  R})\sim T_A({\bf R})$ at small $r$, is violated. To fix
this issue we shall also consider a second possibility to generate
$Q_{s0,{\rm nucleus}}^{2}({\bf R})$:
\begin{equation}   \label{eq:addQs2iniAAMQS}
b)\quad Q_{s0,{\rm nucleus}}^2({\bf R}) = N({\bf R})^{1/\gamma} \,
Q_{s0,\, {\rm proton}}^2~,
\end{equation}
where $\gamma$ is one of the parameters in the initial condition, see
Eq.~(\ref{ic}). This {\it ansatz} --to which we shall refer as {\it modified}-- is motivated by the requirement that
nuclear modification factors should go to unity at large
transverse momentum, $R_{\rm p+Pb}(p_{t}\gg Q_{s})\to 1$, as we shall discuss
in detail later. While we presently lack a solid theoretical
derivation for Eq.~(\ref{eq:addQs2iniAAMQS}) we view it as a
phenomenological way to ensure that the nuclear UGD is additive in the
number of nucleons at small dipole sizes $r$ or high intrinsic
transverse momenta $k_t$.

In both cases, the total number of nucleons that overlap at a given
transverse point, $N({\bf R})$, is calculated via a simple
geometric criterium:
\begin{equation}
N({\bf R}) = \sum\limits_{i=1}^{A} \Theta \left(
\sqrt{\frac{\sigma_0}{\pi}} -  |{\bf R-r_i}|\right)~.
\end{equation}
Some care must be exercised in choosing the transverse area $\sigma_0$
of the large-$x$ partons of a nucleon. $Q_{s0,{\rm proton}}$ corresponds to the
density of large-$x$ sources with $x>x_0$ and should therefore be
energy independent (recoil of the sources is neglected in the
small-$x$ approximation). We therefore take $\sigma_0 \simeq 42$~mb to
be given by the inelastic cross-section at $\surd s =
200$~GeV. However, $\sigma_0$ should not be confused with the energy
dependent inelastic cross section $\sigma_{\rm in}(s)$ of a nucleon
which grows due to the emission of small-$x$ gluons. 

\begin{figure}[htb]
\begin{center}
\includegraphics[width=0.49\textwidth]{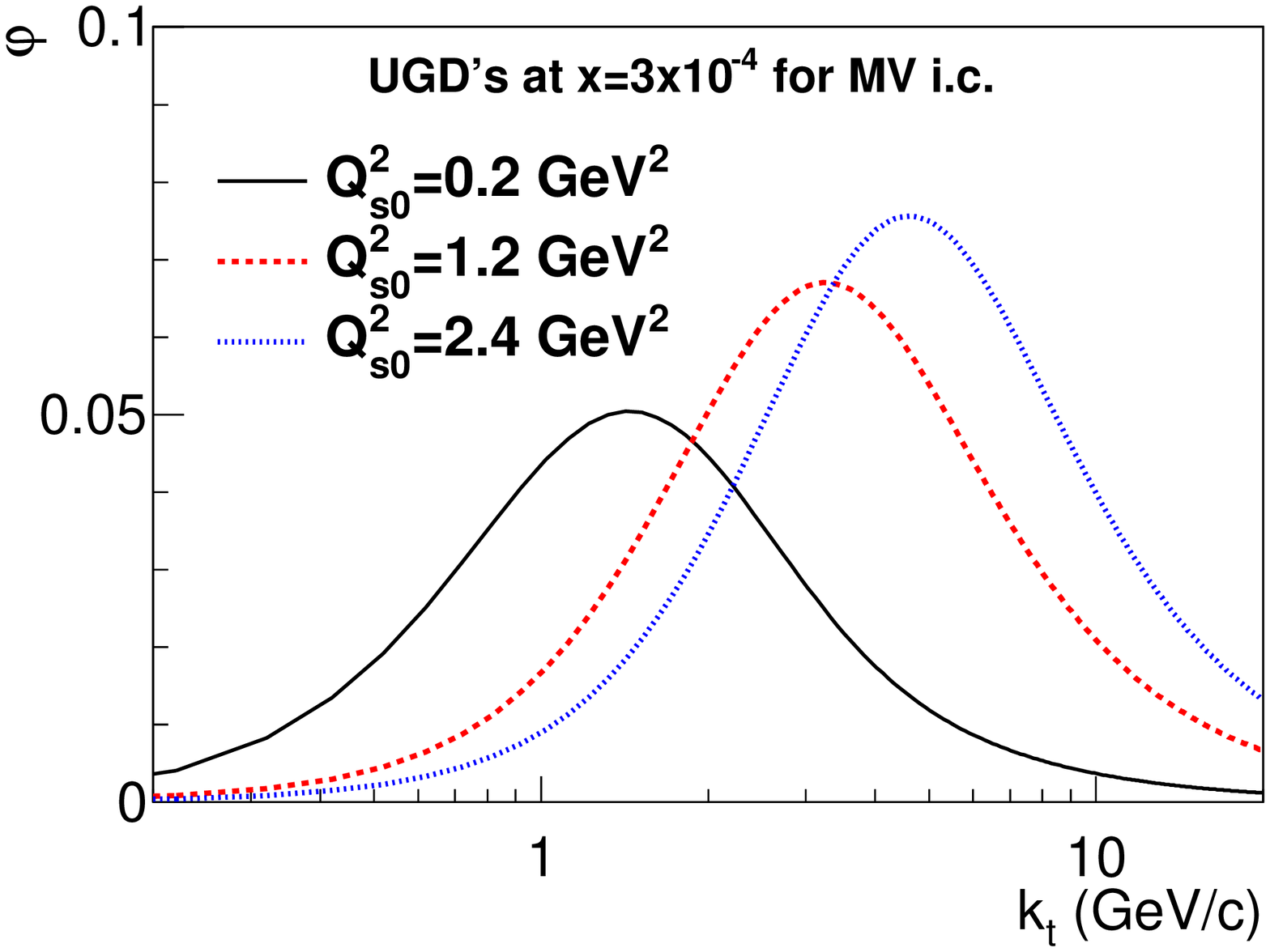}
\includegraphics[width=0.49\textwidth]{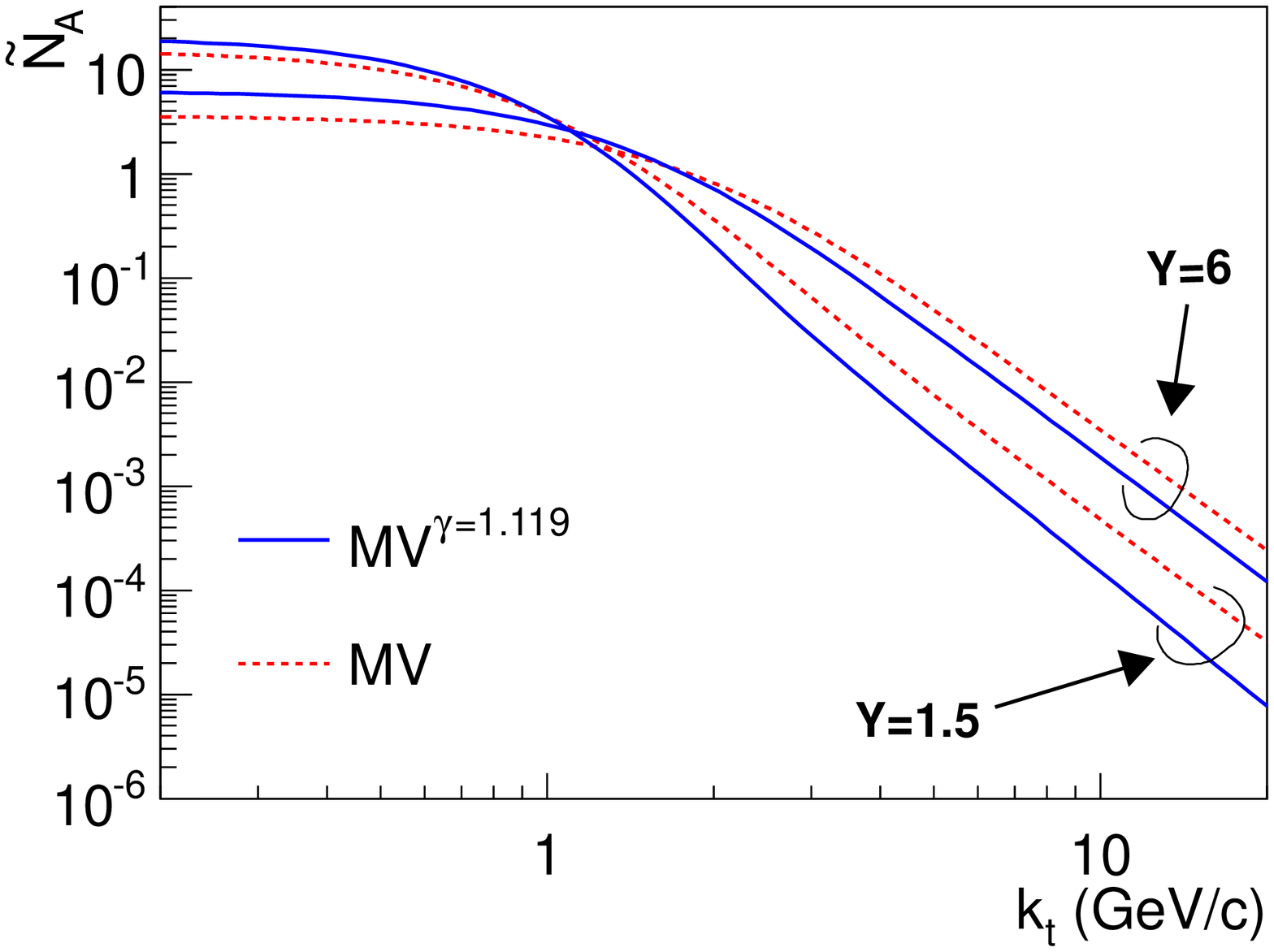}
\end{center}
\vspace*{-.4cm}
\caption[a]{Left: unintegrated gluon distributions $\varphi(k_t;x)$
  for different values of the initial saturation scale evolved to
  $x=3\cdot 10^{-4}$ for MV initial conditions.  Right: Fourier
  transform of the dipole scattering amplitude $\widetilde{\cal
    N}_{A}(k;x)$ for a single nucleon and two different initial
  conditions: MV$^{\gamma=1.119}$ (solid) and MV (dashed) at
  rapidities $Y=\ln(x_0/x)=1.5$ and 6. }
\label{UGDs}
\end{figure}
In Fig.~\ref{UGDs} we plot the UGD 
$\varphi(k_t;x)$
for three different initial saturation scales at $x=3\cdot 10^{-4}$ versus
transverse momentum. The UGD corresponding to a single nucleon peaks
at about $k_t\simeq1$~GeV. The UGDs for larger $Q_{s0}^2$ illustrate
the shift corresponding to a 6-nucleon and 12-nucleon target,
respectively. These curves illustrate why one hopes that particle
production at high energies and/or for heavy targets would become
insensitive to physics at the confinement scale while instead being
dominated by the semi-hard scale generated dynamically.

The plot on the right shows the different slopes for the two
initial conditions (MV$^{\gamma=1.119}$ versus MV). The AAMQS UGD
exhibits a substantial suppression of the high-$k_t$ tail as compared
to MV model initial conditions. At fixed $k_t$ the suppression
diminishes with evolution to higher rapidities. We shall show below
that the AAMQS UGD provides a much better description of semi-hard
$p_t$ spectra in p+p collisions at high energies.

\section{Particle production: $k_t$-factorization and {\it hybrid} formalisms}

In this section we summarize how particle production is calculated
from the UGDs described above. It is useful to first recall some
elementary kinematics. For inclusive production of a single parton
with transverse momentum $p_{t}$ and rapidity $y$, {\em without}
detection of the recoiling particle(s) in the opposite hemisphere, the
$2\to1$ kinematics is such that projectile and target fields are
probed at light cone momenta $x_{1,2}=(p_{t}/\sqrt{s})\, e^{\pm
  y}$. For the upcoming p+Pb run at the LHC we shall assume a
collision energy of $\sqrt{s}=5$~TeV. Hence, the production of
particles with transverse momentum $p_{t}\lesssim 20$ GeV in the
central rapidity region $|y|\lesssim 1$ probe both the projectile and
target wave functions at small values of $x$ below our initial
condition at $x_0=0.01$~\footnote{A more precise estimate should
  consider the kinematic shift induced by the convolution of the
  primary produced parton with the fragmentation function, which is
  indeed taken into account in our calculation later on.}. In this
case we shall use the so-called $k_{t}$-factorization approach. On the
other hand, at more forward rapidities $y\gtrsim 2$, towards the
proton fragmentation region, the momentum carried by the projectile
parton grows large and so we shall resort to the hybrid formalism.

\begin{figure}[tb]
\includegraphics[width=0.49\textwidth]{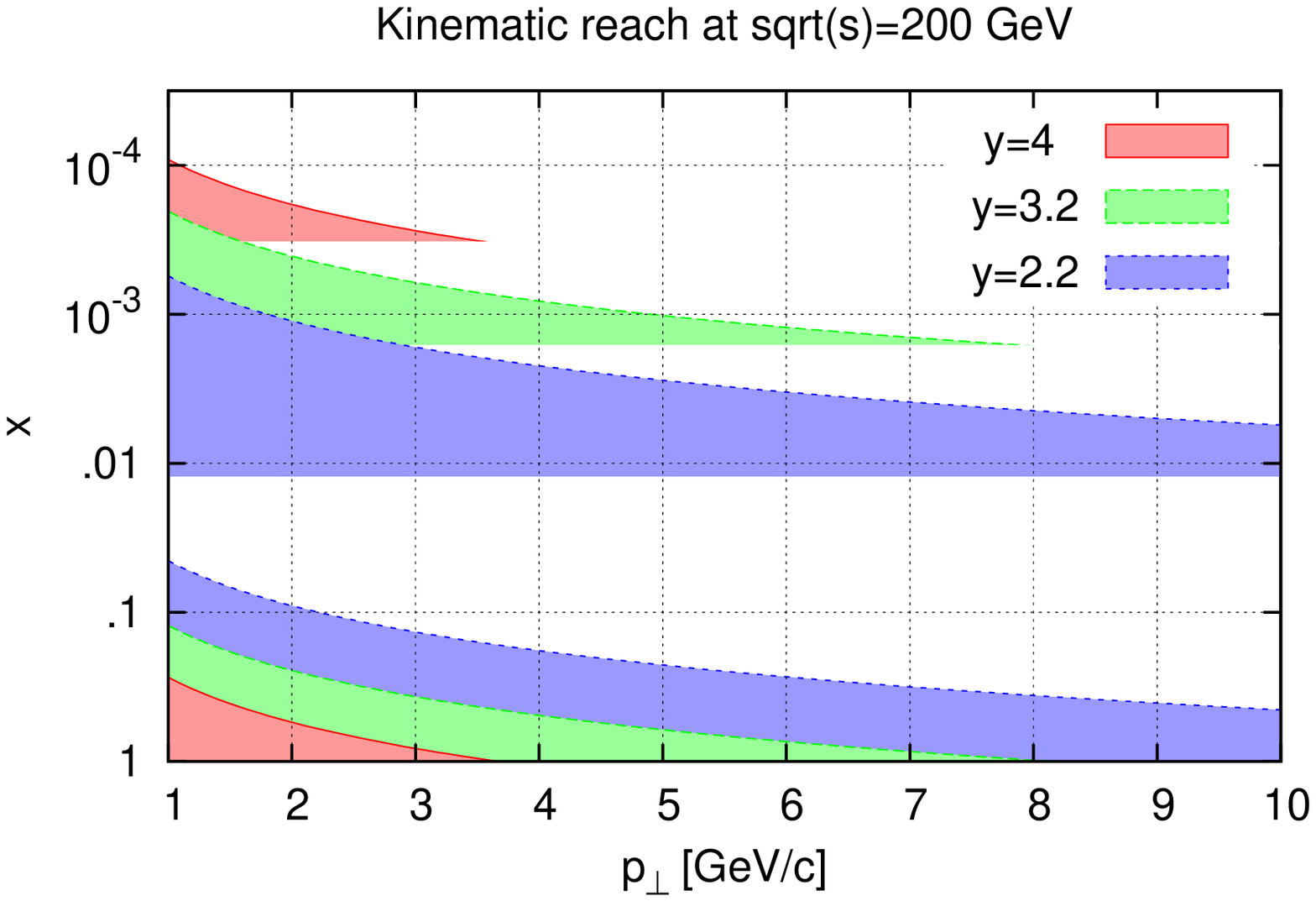}
\includegraphics[width=0.49\textwidth]{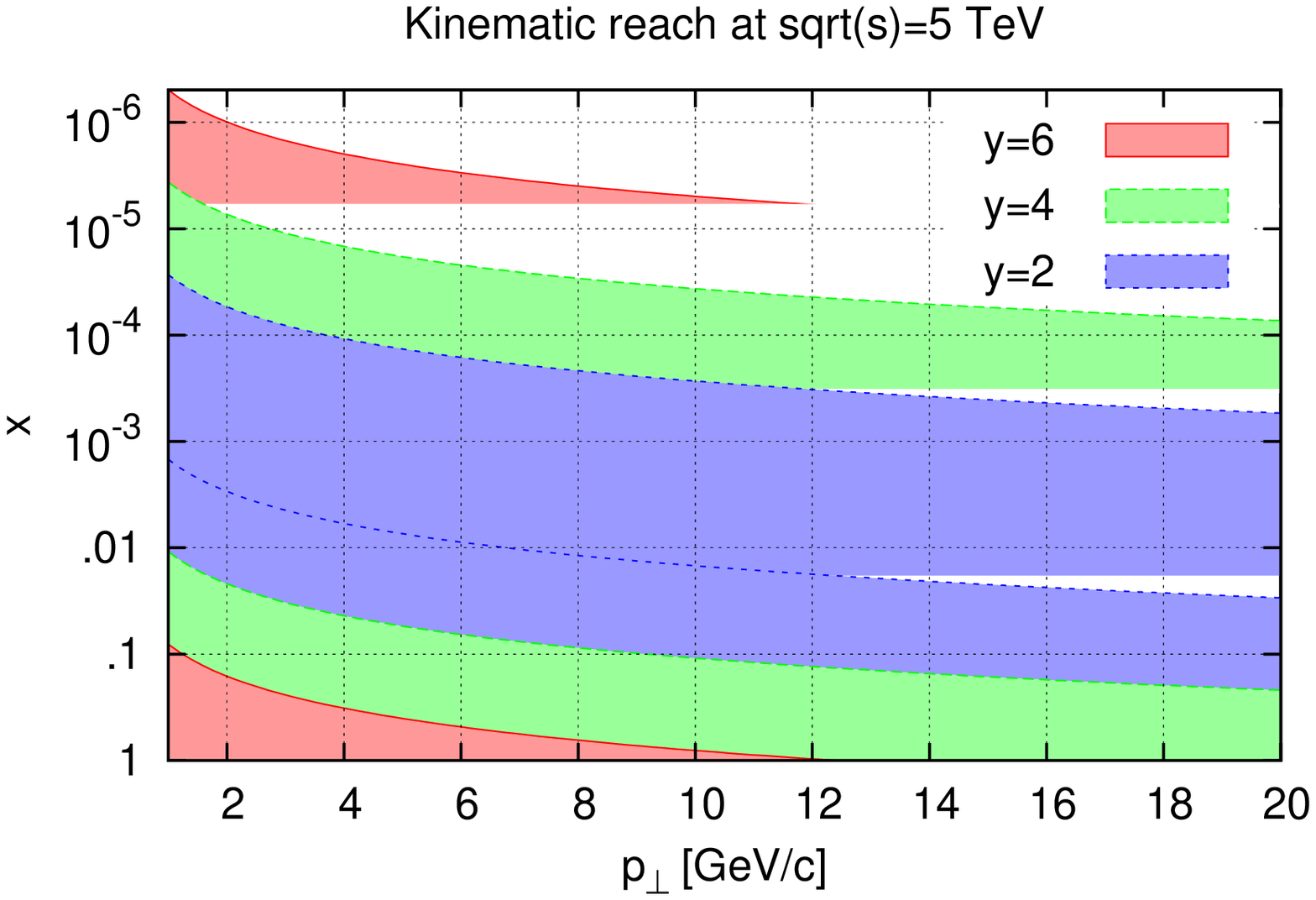}
\caption{Left: $x_{1,2}$ coverage at $y=2.2$, 3.2 and 4.0 at RHIC energy
$\sqrt{s}=200$ GeV.
Hadron $p_\perp$ is limited to $p_T \lsim 7.5$ (3.5) GeV/c at
$y=3.2$ (4.0) at RHIC.
Right: $x_{1,2}$ coverage at $y=4$, 6 at LHC energy.}
\label{fig:x2Cover}
\end{figure}
To outline the limits of applicability of our model let us first
mention that in our terminology ``small-$x$'' refers to $x\le
x_0=0.01$. This is the limit of applicability of the AAMQS fits to
HERA data and coincides with parametric estimates for the validity of
the CGC approach and of coherent interactions. To estimate the range
of $p_{t}$ where our calculations might be valid we again resort to
the AAMQS fits: they start exhibiting some tension with the data when
extended beyond $Q\sim 7\div 10$ GeV. This may indicate failure
of the CGC approach based on rcBK resummation for higher virtualities
or transverse momenta. These limits of applicability are, of course,
merely indicative, since one does not expect a sharp boundary but
rather a smooth transition. The kinematic window accessible at RHIC
and LHC, respectively, is illustrated in Fig.~\ref{fig:x2Cover}.

\subsection{$k_{t}$-factorization}  \label{sec:ktfact}
According to the $k_t$-factorization
formalism~\cite{Kovchegov:2001sc}, the number of gluons produced per
unit rapidity at a transverse position ${\bf R}$ in A+B collisions is
given by
\begin{equation}
\frac{dN^{A+B\to g}}{dy\, d^2p_t\, d^2R} = \frac{1}{\sigma_s}
\frac{d\sigma^{A+B\to g}}{dy\, d^2p_t\, d^2R}\,,
\label{kt}
\end{equation}
where $\sigma_s$ represents the effective interaction area and
$\sigma^{A+B\to g}$ is the cross section for inclusive gluon
production:
\begin{equation}
\frac{d\sigma^{A+B\to g}}{dy\, d^2p_{t}\, d^2R} = K^{k}\,  \frac{2}{C_F} 
\frac{1}{p_t^2} \int^{p_t} \frac{d^2k_t}{4}\int d^2b~
\alpha_s(Q)\,
\varphi_P\left(\frac{|p_t+k_t|}{2},x_1;b\right)\;
\varphi_T\left(\frac{|p_t-k_t|}{2},x_2;R-b\right)~.
\label{kt2}
\end{equation}
$y$ and $p_{t}$ are the rapidity and transverse momentum of the
produced gluon, respectively, while $x_{1,2} = (p_{t} /
\sqrt{s_{NN}})\exp(\pm y)$ and $C_F=(N_c^2-1)/2N_c$. As noted before,
we assume that the local density in each nucleus is homogeneous over
transverse distances of the order of the nucleon radius $R_N$. Thus,
the $b$-integral in~\eq{kt2} yields a geometric factor proportional to
the transverse ``area'' of a nucleon which cancels with a similar
factor implicit in $\sigma_s$ from \eq{kt}, modulo subtleties in the
definition of $\sigma_s$. Note, also, that~(\ref{kt2}) is symmetric
under projectile $\leftrightarrow$ target exchange if, simultaneously,
one lets $y\to -y$.

Eq.~(\ref{kt2}) was written originally for fixed coupling. For
consistency with our running coupling treatment of the small-$x$
evolution of the UGDs, we allow the coupling to run with the momentum
scale. The argument of the running coupling in \eq{kt2} is chosen to
be $Q={\rm max}\{|p_t+k_t|/2,|p_t-k_t|/2\}$.  However, we found rather
weak sensitivity to the particular choice of scale because
$\varphi\to0$ as $k_t\to0$ due to the saturation of $\mathcal{N}(r)$
at large dipole sizes $r$, see above. In principle one could improve
on this educated guess by using the ``running coupling
$k_t$-factorization'' formula derived recently in
ref.~\cite{Horowitz:2010yg}. Most importantly, the $x$-dependence of
the dipole scattering amplitude obtained by solving the rcBK equation
encodes all the collision energy and rapidity dependence of the gluon
production formula \eq{kt2}.

The normalization factor $K^{k}$ introduced in the $k_t$-factorization
formula~(\ref{kt2}) above lumps together higher-order corrections,
sea-quark contributions and, effectively, other dynamical effects not included in the CGC formulation. It can be fixed approximately from
the charged particle transverse momentum distribution in p+p
collisions at 7~TeV, see below. Although its precise value depends
on the UGD and on the fragmentation function, we typically find
$K^{k}\simeq 1.5-3$, which appears reasonable.

Eq.~(\ref{kt2}) is the starting point for all observables discussed below. In
particular, the charged particle multiplicity and the transverse
energy can be obtained by integrating over the transverse plane and
$p_t$,
\begin{eqnarray}
\frac{dN_{\rm ch}}{dy} &=& \frac{2}{3} \kappa_g \int d^2R \int d^2p_t \,
     \frac{dN^{A+B\to g}}{dy\, d^2p_t\, d^2R} ~, \label{eq:totdNdy}\\
\frac{dE_t}{dy} &=& \int d^2R \int d^2p_t \, p_t\,
     \frac{dN^{A+B\to g}}{dy\, d^2p_t\, d^2R} ~. \label{eq:totdEtdy}
\end{eqnarray}
Note that a low-$p_t$ cutoff is not required since the integration
over $k_t$ in~(\ref{kt2}) extends only up to $p_t$. The saturation of
the gluon distribution functions guarantees that the dominant scale in
the transverse momentum integrations is the saturation
momentum. Similar to other CGC-based approaches, our
Eq.~(\ref{eq:totdNdy}) assumes that the total hadron multiplicity is
proportional to the initial gluon multiplicity through an (energy and
centrality independent) {\it gluon multiplication factor}
$\kappa_g$ \footnote{This factor does not enter
  Eq.~(\ref{eq:totdEtdy}) for the transverse energy because we assume
  that final-state gluon showering and hadronization conserves the
  energy per unit rapidity.}. In order to reproduce both RHIC and LHC data
on charged hadron multiplicities in heavy-ion collisions we fix it to
be $\kappa_g\simeq5$; small adjustments of this normalization factor
may be required for p+p and p+Pb collisions as discussed below. This
could be due to the fact that we here assume local rcBK evolution
without explicit impact parameter dependence; further, due to our
ignorance about how to hadronize the small-$x$ gluons (which also
determines the $y\to\eta$ Jacobian given below) etc. The precise
value of $\kappa_g$ does of course depend on the value of the
$K$-factor which has been fixed independently since $dN/dy$ only
involves their product.

The upper limit in the integrals over the gluon transverse momentum in
$dN_{\rm ch}/dy$ and $dE_t/dy$ has been taken as $p_t^{\rm
  max}=12$~GeV; if the integrals are extended further then a slight
adjustment of the normalization factors $\kappa_g$ and $K$ may be
required. In order to compare our results for initial gluon
production to the final state distributions of charged particles one
has to translate the rapidity distributions into pseudo-rapidity
distributions through the $y\to\eta$ Jacobian
\beq
\frac{dN_{\rm ch}}{d\eta} = \frac{\cosh\eta}
{\sqrt{\cosh^2\eta + m^2/P^2}}\, \frac{dN_{\rm ch}}{dy}~~~~,~~~~
\frac{dE_t}{d\eta} = \frac{\cosh\eta}
{\sqrt{\cosh^2\eta + m^2/P^2}}\, \frac{dE_t}{dy}~,
\eeq
with $y=\frac{1}{2}\ln\, (\sqrt{\cosh^2\eta + m^2/P^2}+\sinh\eta)/
(\sqrt{\cosh^2\eta + m^2/P^2}-\sinh\eta)$. We
assume that in this Jacobian $m=350$~MeV and $P=0.13~\mathrm{GeV} + 
0.32~\mathrm{GeV} \, (\sqrt{s}/1~\mathrm{TeV})^{\, 0.115}$ which leads
to a reasonably good description of the pseudo-rapidity distribution
of charged particles in p+p collisions at LHC energies, see below.

In turn, the single inclusive spectra at perturbatively large
transverse momenta can be obtained by folding Eq.~(\ref{kt2}) for
gluon production with the corresponding gluon fragmentation function:
\begin{equation}
 \frac{dN^{A+B\to hX}}{dy\, d^2p_t}=\int d^2R \int
 \frac{dz}{z^2} \, D_g^h\left(z=\frac{p_t}{k_t},Q\right) \,
     \frac{dN^{A+B\to g}}{dy\, d^2q_t\, d^2R} \label{eq:FFconv}\,.
\end{equation}
In~(\ref{eq:FFconv}) the integral over the hadron momentum
fraction is restricted to $z\ge0.05$ to avoid a violation of the
momentum sum rule. The scale dependence of the FF of course emerges
from a resummation of collinear singularities via the DGLAP equations
and so its use in the $k_\perp$-factorization formula is not entirely
justified.

\subsection{Hybrid formalism}

Moving away from central rapidity towards the projectile fragmentation
region its wave function is probed at larger and larger momentum
fraction $x_1$ which will eventually exceed $x=0.01$. In this case the
so-called hybrid formalism~\cite{Dumitru:2005gt} is better suited for
particle production. We shall employ the following expression for the
differential cross section for production of a hadron with transverse
momentum $k$ and pseudorapidity\footnote{At forward rapidities the
  distinction between $\eta$ and $y$ becomes less relevant.} $\eta$:
\begin{equation}
\frac{dN^{pA\to
    hX}}{d\eta\,d^2k}=K^{h}\left(\left[\frac{dN_h}{d\eta\,d^2k}\right]_{{\rm
    el}}+\left[\frac{dN_h}{d\eta\,d^2k}\right]_{{\rm inel}} \right)
\label{hyb}
\end{equation}
where the subscripts {\it el} and {\it inel} stand for elastic and
inelastic contributions\footnote{As a NLO contribution, the latter
  need not be positive definite, see below.}, respectively. We again
allow for the presence of a K-factor, $K^h$, to absorb higher order
corrections. The first term in Eq.\ (\ref{hyb}), the elastic
contribution, is given by~\cite{Dumitru:2005gt}
\begin{eqnarray}
\left[\frac{dN_h}{d\eta\,d^2k}\right]_{{\rm el}} = \frac{1}{(2\pi)^2\!\!}
\int_{x_F}^1\frac{dz}{z^2}\, \left[\sum_{q}x_1f_{q/p}
(x_1,Q^{2})\; \tilde{N}_F\left(x_2,\frac{p_t}{z}\right) \; D_{h/q}(z,Q^{2})
  \right. \nonumber\\ +\left. 
x_1f_{g/p}(x_1,Q^{2})\; \tilde{N}_A \left(x_2,\frac{p_t}{z}\right)\;
D_{h/g}(z,Q^{2}) \right] 
\label{hybel}\,,
\end{eqnarray}
and corresponds to scattering of collinear partons from the projectile
on the target. The $2\to 1$ kinematics sets
$x_{1,2}=(p_t/z\sqrt{s_{NN}})\, \exp(\pm y)$ and
$x_F\simeq (p_t/\sqrt{s_{NN}})\, \exp\, \eta$.
The projectile is described by standard collinear parton distribution
functions (PDFs) but its partons acquire a large transverse momentum
$k$ due to (multiple) scattering from the small-$x$ fields of the
nucleus which are described by the corresponding UGDs in the adjoint
or fundamental representation $\tilde{N}_{A(F)}$, see
Eqs.~(\ref{phihyb}). The hadronization of the scattered parton into a
hadron is described by the usual fragmentation function (FF) of
collinear factorization, $D_{h/j}$. Both the PDF and the FF are
evaluated at the factorization scale $Q$. We shall explore the
sensitivity to the choice of factorization scale by letting it vary
within the range $Q=(k/2,2k)$.

The inelastic term in Eq.~(\ref{hyb}) has been calculated recently in
ref.~\cite{Altinoluk:2011qy}. It reads
\begin{eqnarray}
\left[\frac{dN_h}{d\eta\,d^2k}\right]_{{\rm inel}}=
\frac{\alpha_{s}(Q)}{2\pi^{2}} \int_{x_F}^1\frac{dz}{z^2}\frac{z^4}{k^4}
\int^{Q}\!\frac{d^2q}{(2\pi)^2}q^{2} \tilde{N}_F(x_2,q)x_1\!
\int_{x_1}^{1} \frac{d\xi}{\xi}\,\!\!\! \sum_{i,j=q,\bar{q},g}\!\!\!
w_{i/j}(\xi) P_{i/j}(\xi) f_{j}(\frac{x_{1}}{\xi},Q^{2}) D_{h/j}(z,Q^{2})\,,
\label{hybinel}
\end{eqnarray}
where $P_{i/j}$ are the LO DGLAP splitting functions for the different
parton species $i,j=q,\bar{q},g$. Note that endpoint singularities for
$q\to q$ and $g\to g$ splitting are regulated via the usual ``+
prescription''; therefore, the contribution from Eq.~(\ref{hybinel})
is actually negative in parts of phase space. Explicit expressions for
the weight functions $w_{i/j}(\xi)$ are given in Eqs.~(74-77) of
ref~\cite{Altinoluk:2011qy} and shall not be repeated here.

The inelastic term corresponds to an alternative channel for hard
production: partons with high transverse momentum can occur in the
wave function of the incoming proton due to large-angle
radiation. Those may then scatter off the target with only a small
momentum transfer to finally fragment into a high-$p_t$ hadron.  It
should be noted that the inelastic term accounts for part of the full
NLO corrections to the hybrid formalism. A calculation of the full NLO
corrections to the hybrid formalism has been recently presented in
\cite{Chirilli:2011km,Chirilli:2012jd}. While a numerical
implementation of the full NLO corrections would be necessary, such
task is beyond the scope of this paper and we leave it for future
work. The evaluation of the inelastic term in this work provides an
estimate of the numerical importance of the full NLO corrections.

The inelastic contribution involves an additional power of the
coupling $\alpha_{s}$ but also comes with a factor $\sim \log(
k^2/Q_{s,T}^2)$~\cite{Altinoluk:2011qy} and so is expected to be
significant at high transverse momenta and not too forward rapidities,
far from the kinematic boundary.  These expectations have been first
verified numerically in ref.~\cite{JalilianMarian:2011dt}. Finally, we
note that the scale for the running of the coupling in
Eq.~(\ref{hybinel}) cannot be determined from the calculation of
ref.~\cite{Altinoluk:2011qy} --a full NNLO calculation would be
necessary to determine the scale for the running of the coupling at
NLO-- and should be considered as a free parameter. In order to asses
the uncertainty related to the choice of scale for the coupling in the
inelastic term, we shall consider two possibilities: assuming a
constant value $\alpha_{s}=0.1$ (similar to what was done in an
earlier study~\cite{JalilianMarian:2011dt}) or, alternatively,
assuming one-loop running at the factorization scale $Q$ by replacing
$\alpha_{s}\to\alpha_{s}(Q)$ in Eq.~(\ref{hybinel}).

\section{The baseline: proton+proton collisions}

In this section we first present our results for proton-proton
collisions. We restrict to CM energies in the TeV regime so that the
typical parton momentum fractions involved in semi-hard hadron
production remain small. A good description of $p_\perp$ distributions
in p+p is of course required for a reliable calculation of the spectra
in minimum-bias p+A collisions as well as for the $R_{pA}$ nuclear
modification factor. Furthermore, it is important to check consistency
of the UGD in DIS and hadronic collisions.  In what follows we shall
use three different sets of fragmentation functions:
LO-KKP~\cite{Kniehl:2000fe} and
DSS~\cite{deFlorian:2007aj,deFlorian:2007hc} at LO and NLO,
respectively.

\begin{figure}[htb]
\begin{center}
\includegraphics[width=0.49\textwidth]{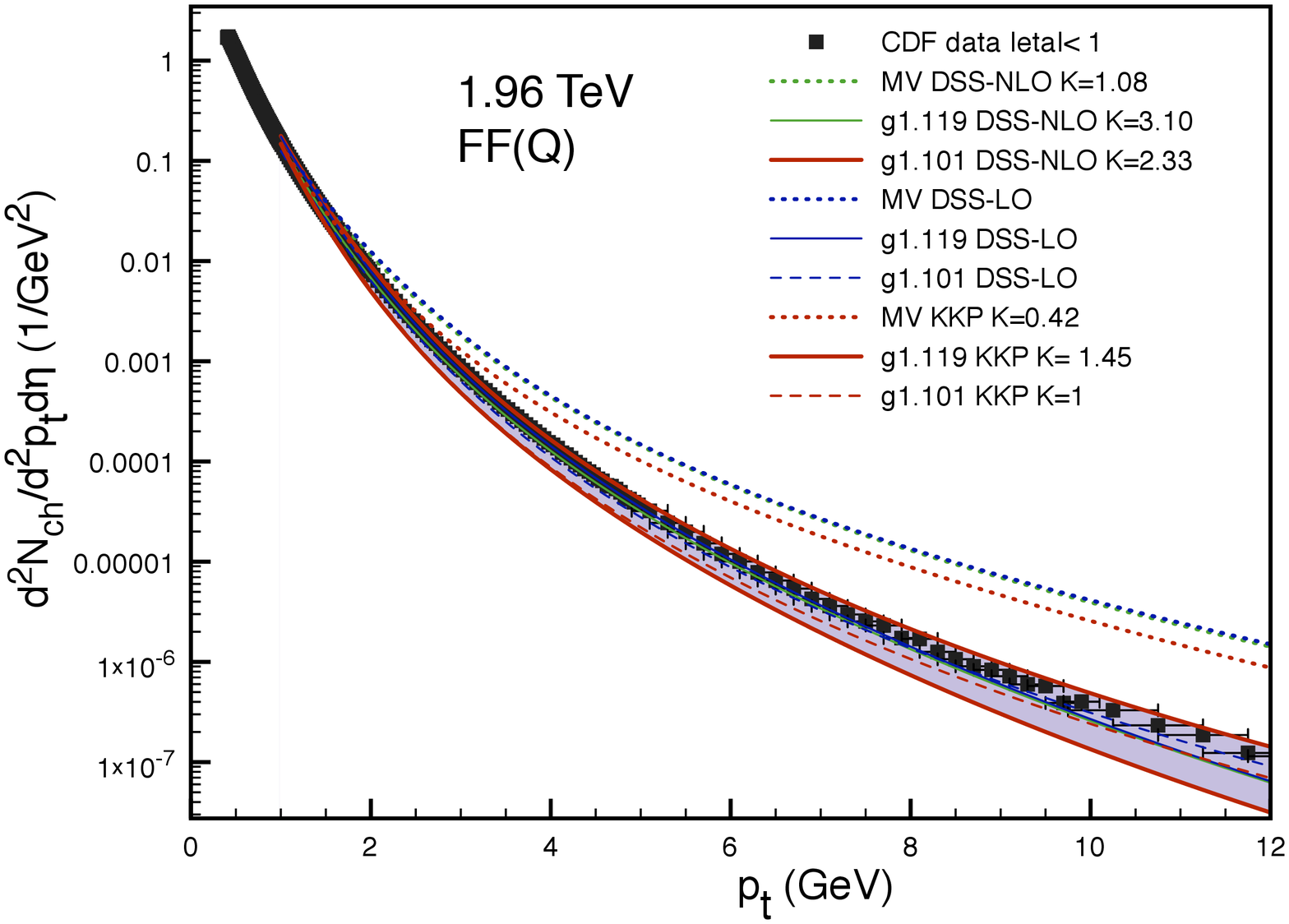}
\includegraphics[width=0.49\textwidth]{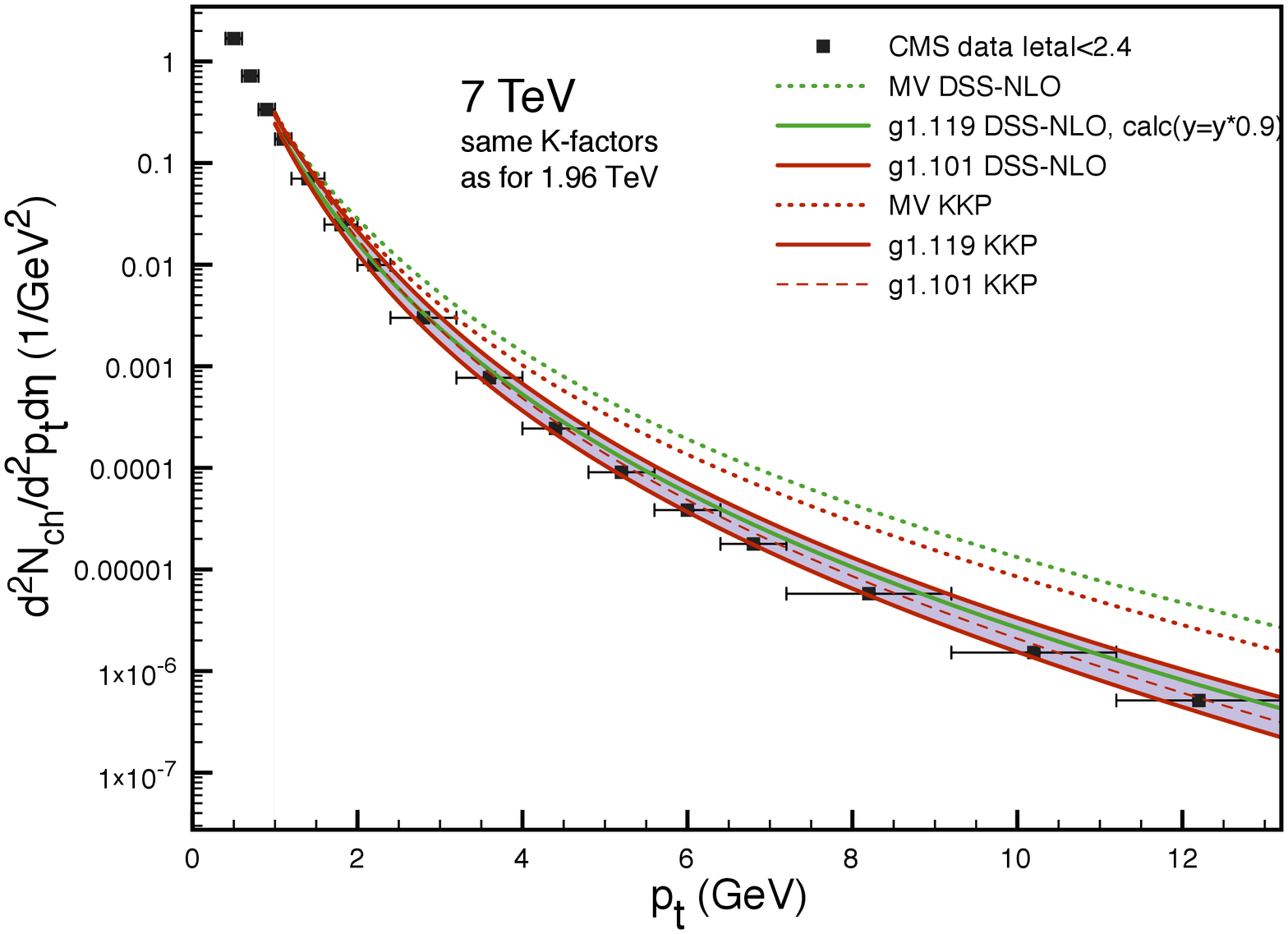}
\end{center}
\vspace*{-0.5cm}
\caption[a]{Transverse momentum distribution of charged particles in
  the central region of p+$\overline{\rm p}$ collisions at $\surd s =
  1.96$~TeV (left), and p+p collisions at $\surd s = 7$~TeV (right).
  CDF and CMS data from refs.~\protect\cite{Aaltonen:2009ne}
  and~\protect\cite{Khachatryan:2010us}, respectively.}
\label{fig:ppspect}
\end{figure}
In Fig.~\ref{fig:ppspect} we show the transverse momentum
distributions of charged particles in the regime of semi-hard $p_t$
for inclusive p+p collisions at $\surd s = 1.96$~TeV and 7~TeV,
respectively. We have tested various combinations of UGDs and FFs and
in each case matched the $K$-factor to the data at $p_\perp=1$~GeV.

First, we note that the UGD sets with the steeper initial gluon
spectrum (UGD sets g1119 and g1101) lead to significantly better
agreement with the data than the classical ``MV model'' initial
condition with $\gamma=1$; the latter leads to a $p_\perp$ spectrum
far outside the experimental error bars. This establishes consistency
within our framework of the UGDs with DIS and hadron-hadron
collisions, since those two sets provide a much better
$\chi^{2}/d.o.f$ in fits to e+p data than the MV one. Recall, also,
that in hadronic collisions $x\sim p_\perp$ for spectra at fixed
rapidity. Hence, the shape of the spectra does provide a direct test
of the rcBK evolution speed.

\begin{figure}[htb]
\begin{center}
\includegraphics[width=0.45\textwidth]{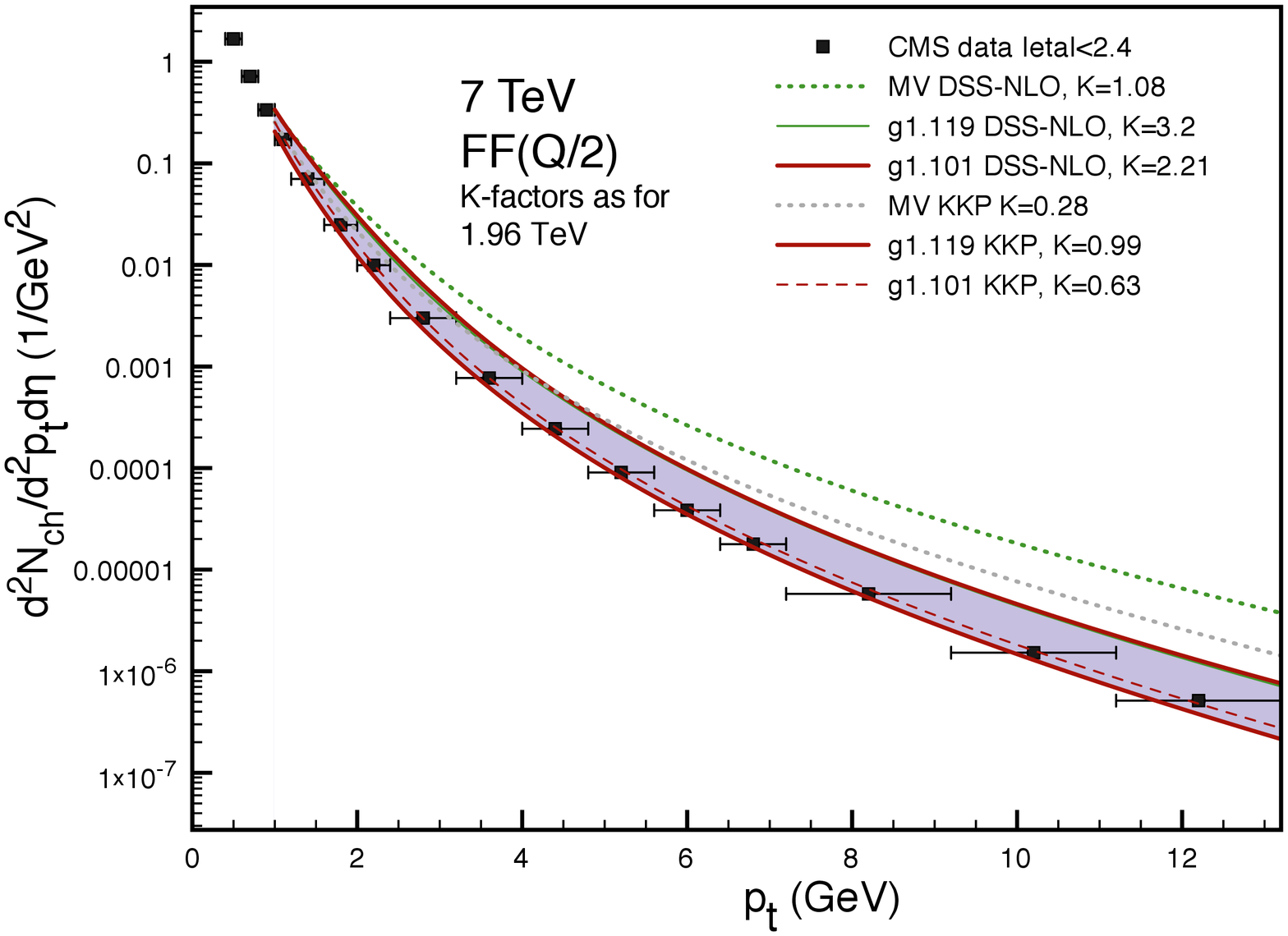}
\includegraphics[width=0.45\textwidth]{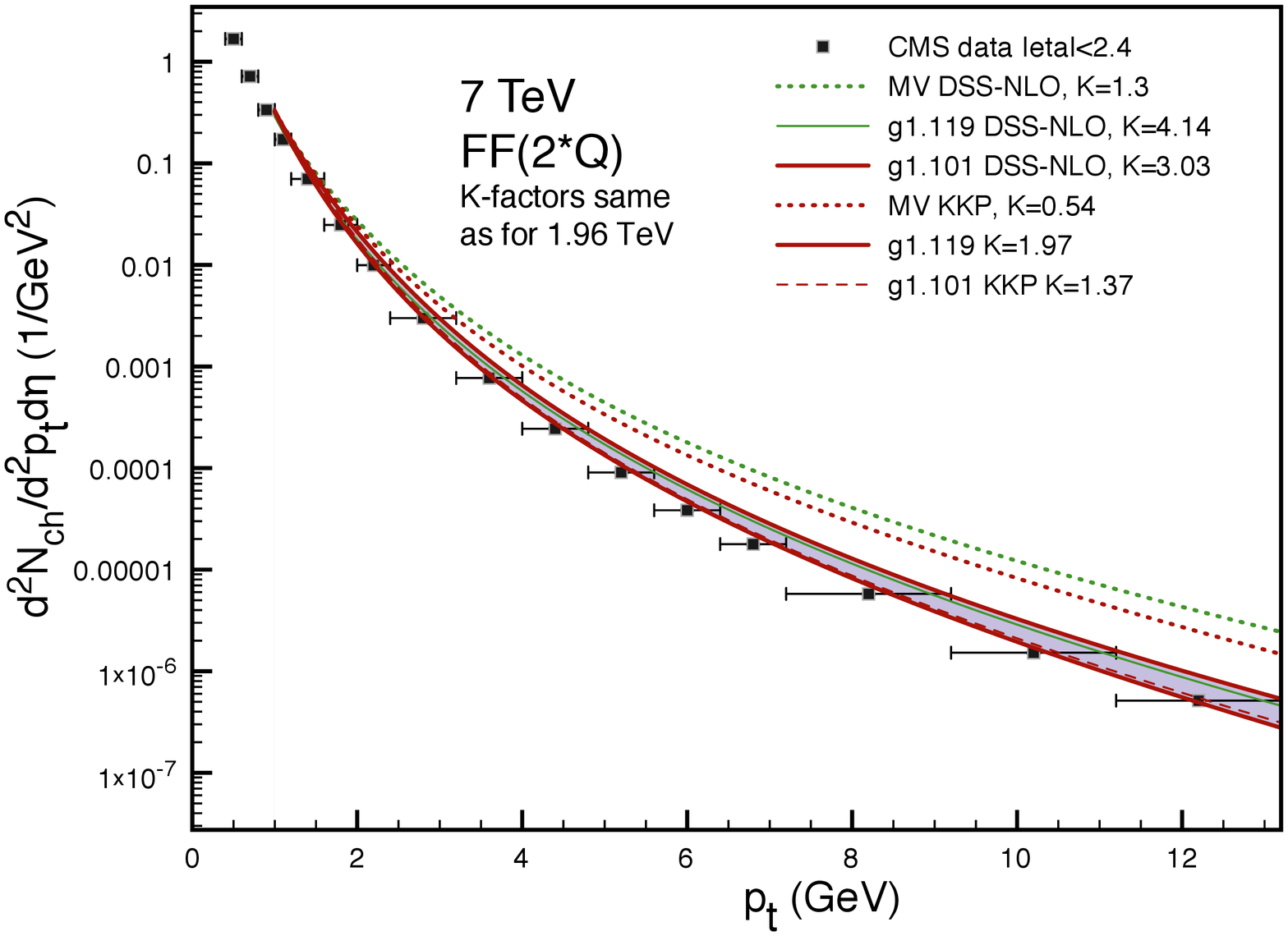}
\end{center}
\vspace*{-0.5cm}
\caption[a]{Transverse momentum distribution of charged particles in
  the central region of p+p collisions at $\surd s = 7$~TeV. The scale
  in the FF is taken to be $Q=p_\perp/2$ (left) or $Q=2p_\perp$
  (right), respectively.}
\label{fig:ppspectQ}
\end{figure}
In Fig.~\ref{fig:ppspectQ} we check that changing the scale in the FF
from $Q=p_\perp/2$ to $Q=p_\perp$ to $Q=2p_\perp$ is mainly absorbed
into a redefinition of the $K$-factor; in what follows we shall fix
$Q=p_\perp$. Similarly, switching from DSS-LO to DSS-NLO FFs
essentially leads to identical spectra once the $K$-factor is
adjusted; see Fig.~\ref{fig:ppspect} (left). On the other hand, we
obtain slightly harder spectra with DSS versus KKP fragmentation
functions, as expected.

\begin{figure}[htb]
\begin{center}
\includegraphics[width=0.45\textwidth]{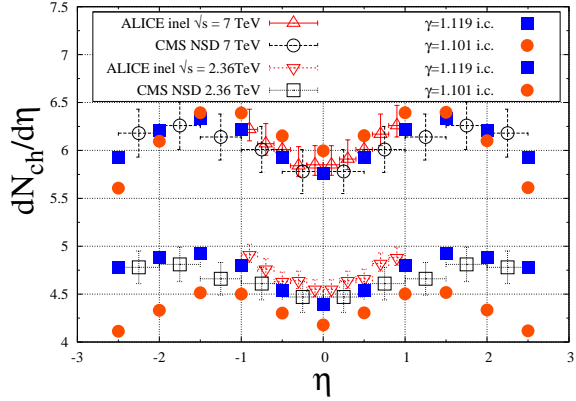}
\end{center}
\vspace*{-.4cm}
\caption[a]{Charged particle multiplicity as a function of
  pseudorapidity for p+p collisions at $\surd s = 2.36$~TeV and 7~TeV,
  respectively, for two different UGD sets (g1119 and g1101). ALICE and CMS data from
  refs.~\cite{Aamodt:2010ft,Aamodt:2010pp,Khachatryan:2010xs,Khachatryan:2010nk}.}
\label{fig:ppmult}
\end{figure}
We now turn to the pseudo-rapidity dependence of $p_\perp$-integrated
multiplicities. As already mentioned above, here we can not convolute
the gluon spectrum with a fragmentation function. Instead, our crude
``hadronization model'' consist in assuming that on average over
events the number of charged particles is proportional to the number
of initially produced gluons. The number $\kappa_g$ of final hadrons
per initial gluon is a free parameter. Its precise value will depend
on the UGD, on the assumed impact parameter distribution of valence
charges in the nucleon, on the $\partial y/\partial\eta$ Jacobian and
so on.

In Fig.~\ref{fig:ppmult} we compare to ALICE and CMS data from
refs.~\cite{Aamodt:2010ft,Aamodt:2010pp,Khachatryan:2010xs,Khachatryan:2010nk}.
For UGD 111 we have adjusted the normalization relative to that
determined from Pb+Pb collisions by a factor of 1.24; the energy and
rapidity dependence of $dN_{\rm ch}/d\eta$ is consistent with the
data. On the other hand, UGD 115 does not require any particular
adjustment of normalization but appears to predict a slightly too steep
energy dependence of the multiplicity and a narrower rapidity
distribution (given our specific $\partial y/\partial\eta$ Jacobian).

\section{Multiplicity in p+Pb collisions}

\begin{figure}[htb]
\begin{center}
\includegraphics[width=0.45\textwidth]{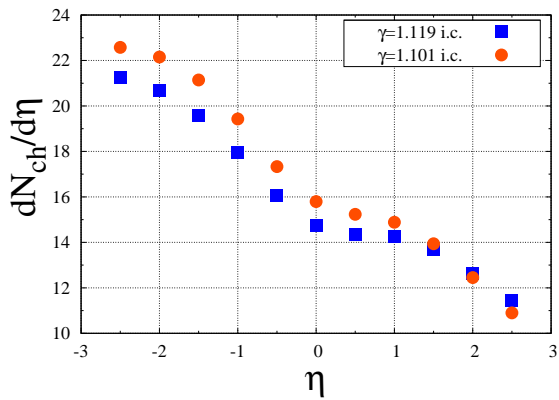}
\end{center}
\vspace*{-.4cm}
\caption[a]{Predicted pseudo-rapidity distribution of charged
  particles in minimum-bias p+Pb collisions at $\surd s=5$~TeV for
  two different UGDs.}
\label{fig:pPbmult}
\end{figure}
In Fig.~\ref{fig:pPbmult} we present our predictions for
$p_\perp$-integrated multiplicities in minimum bias p+Pb
collisions. According to our findings from above, for UGD set g1119 we
use the exact same normalization as for Pb+Pb and p+p collisions. On
the other hand, we had found that the UGD set g1101 requires a
correction of +24\% to reproduce the measured multiplicity in p+p
collisions at 2360 and 7000~GeV. Accordingly, for this UGD set and
min.~bias p+Pb collisions we have increased by hand the normalization
by +10\% (relative to Pb+Pb).  A prediction for the charged
multiplicity for the UGD with MV-model i.c.\ ($\gamma=1$) can be found
in ref.~\cite{Albacete:2010ad}. We mention that the current results
are similar to other predictions based on the idea of gluon
saturation~\cite{Tribedy:2011aa,Rezaeian:2011ia,Dumitru:2011wq}. This
illustrates that indeed the energy and system size dependence of the
multiplicity is governed mainly by the dependence of the saturation scale
$Q_s$ on the target thickness and on $x$.

\section{Single inclusive spectra in p+Pb collisions}

In this section we present single-inclusive charged hadron transverse
momentum distributions for p+Pb collisions at $\surd s=5$~TeV. For
pseudo-rapidities near the central region, $|\eta|<2$, the relevant
light-cone momentum fractions in both projectile and target are
comparable and small and so we use $k_\perp$ factorization,
section~\ref{sec:ktfact}. We fix the $K$-factor to the value extracted
from p+p collisions as described in the previous section. Our default
fragmentation function is KKP-LO evaluated at the scale $Q^2 = k_t^2$.

We shall also show the ``nuclear modification factor'' $R_{\rm p+Pb}$
defined as
\beq  \label{eq:RpPb}
R_{\rm p+Pb}(p_\perp) \equiv \frac{1}{\langle N_{\rm coll}\rangle}\; 
\frac{dN_{\rm ch}^{\rm p+Pb}/d\eta\, d^2p_\perp} {dN_{\rm ch}^{\rm
    p+p}/d\eta\, d^2p_\perp}~,
\eeq
where $\langle N_{\rm coll}\rangle$ denotes the mean number of
binary nucleon-nucleon collisions in a given centrality class; it is
obtained from a standard MC Glauber model using an inelastic cross
section of $\sigma_{\rm in}(\surd s=5\,\mathrm{TeV}) = 67$~mb. For
minimum bias collisions this leads to $\langle N_{\rm coll}
\rangle\approx7$.

To facilitate interpretation of the numerical results we shall not
restrict to the AAMQS-like UGDs with $\gamma>1$ initial condition but
also show some curves obtained with the UGD with $\gamma=1$ MV-model
initial condition. We stress that although this UGD does not provide a
good description of neither DIS data on protons nor of semi-hard $p_t$
spectra in p+p collisions, it has not been directly tested against
nuclear data yet and, therefore, remains a viable candidate for the
initial condition for the evolution of nuclear wave functions.

\begin{figure}[htb]
\begin{center}
\includegraphics[width=0.48\textwidth]{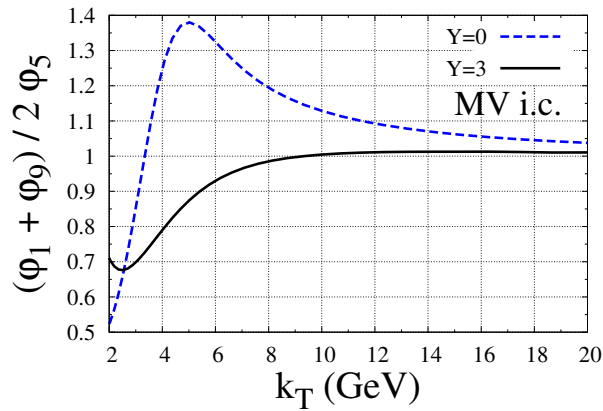}
\end{center}
\vspace*{-0.4cm}
\caption[a]{Average of the gluon densities of a single nucleon and a
  nine nucleon target divided by the gluon density of a five nucleon
  target, for $Y=0$ (MV i.c.) and $Y=3$ (MV + rcBK), respectively. }
\label{fig:UGDratio}
\end{figure}
We first illustrate the effects of quantum evolution and of
fluctuations in the thickness of the target. Fig.~\ref{fig:UGDratio}
compares the average UGD of a 1-nucleon and 9-nucleon target to that
of a 5-nucleon target. The MV initial condition at $Y=0$ shows a
strong suppression of this ratio at low intrinsic transverse momentum
followed by a ``Cronin-like'' peak and an asymptotic approach to 1
from above (the leading higher-twist correction in the MV model is
positive, $\sim +Q_s^4/k_T^6$; appendix~B in
ref.~\cite{Gelis:2001da}). Neglecting fluctuations in the number of
target nucleons could clearly distort the resulting $R_{\rm pA}$ ratio
significantly. Actually, this could be origin of the slight
differences between our predictions for $R_{pPb}$ and those presented
in~\cite{Tribedy:2011aa} (shown in Fig.~\ref{fig:RpPbmb}
below) with similar dynamical input but where the nuclear
geometry is treated in a mean field approach, hence neglecting
fluctuations. Resummation of small-$x$ quantum fluctuations removes
the enhancement at intermediate $k_T$.

\begin{figure}[htb]
\begin{center}
\includegraphics[width=0.48\textwidth]{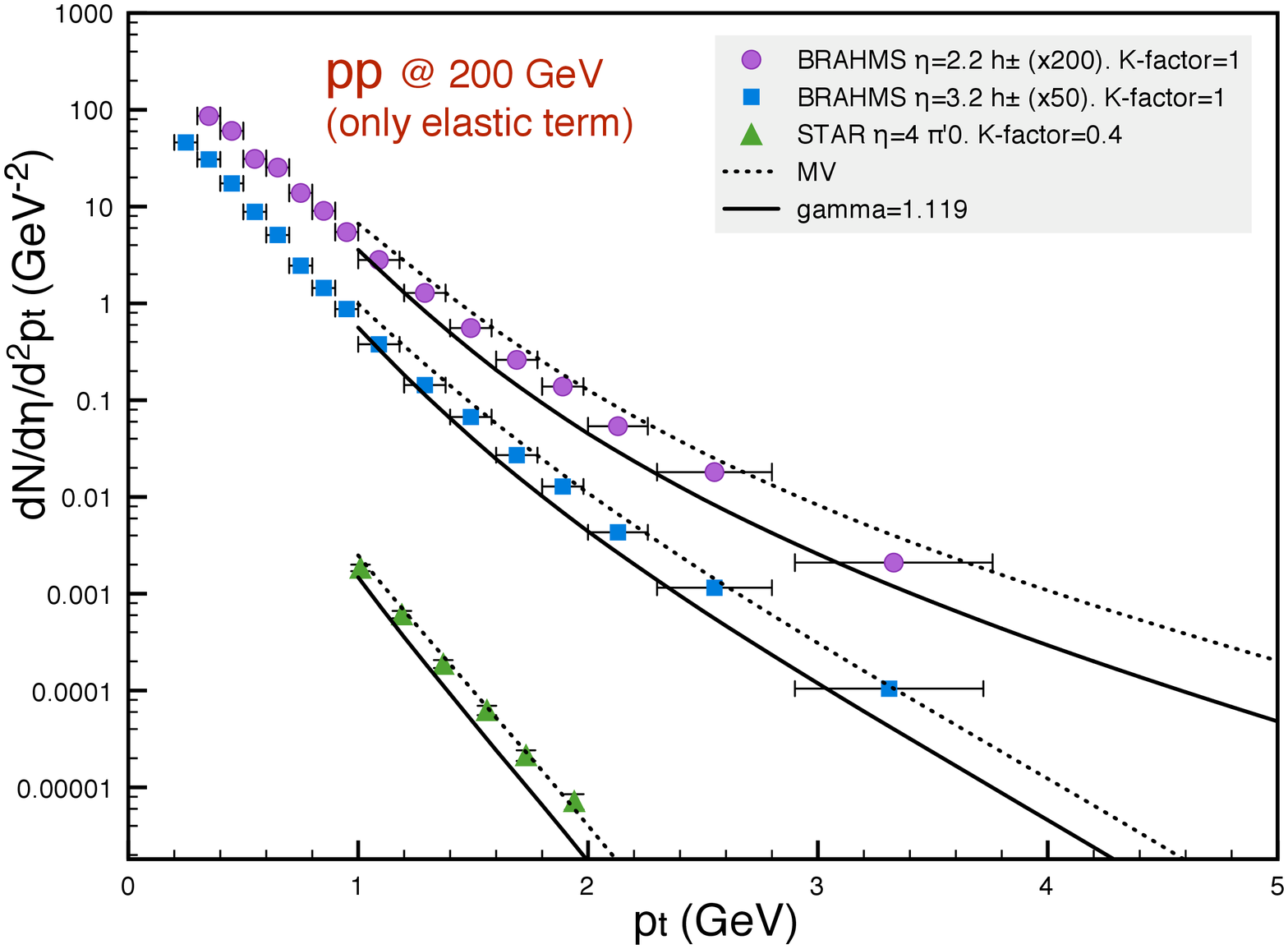}
\includegraphics[width=0.48\textwidth]{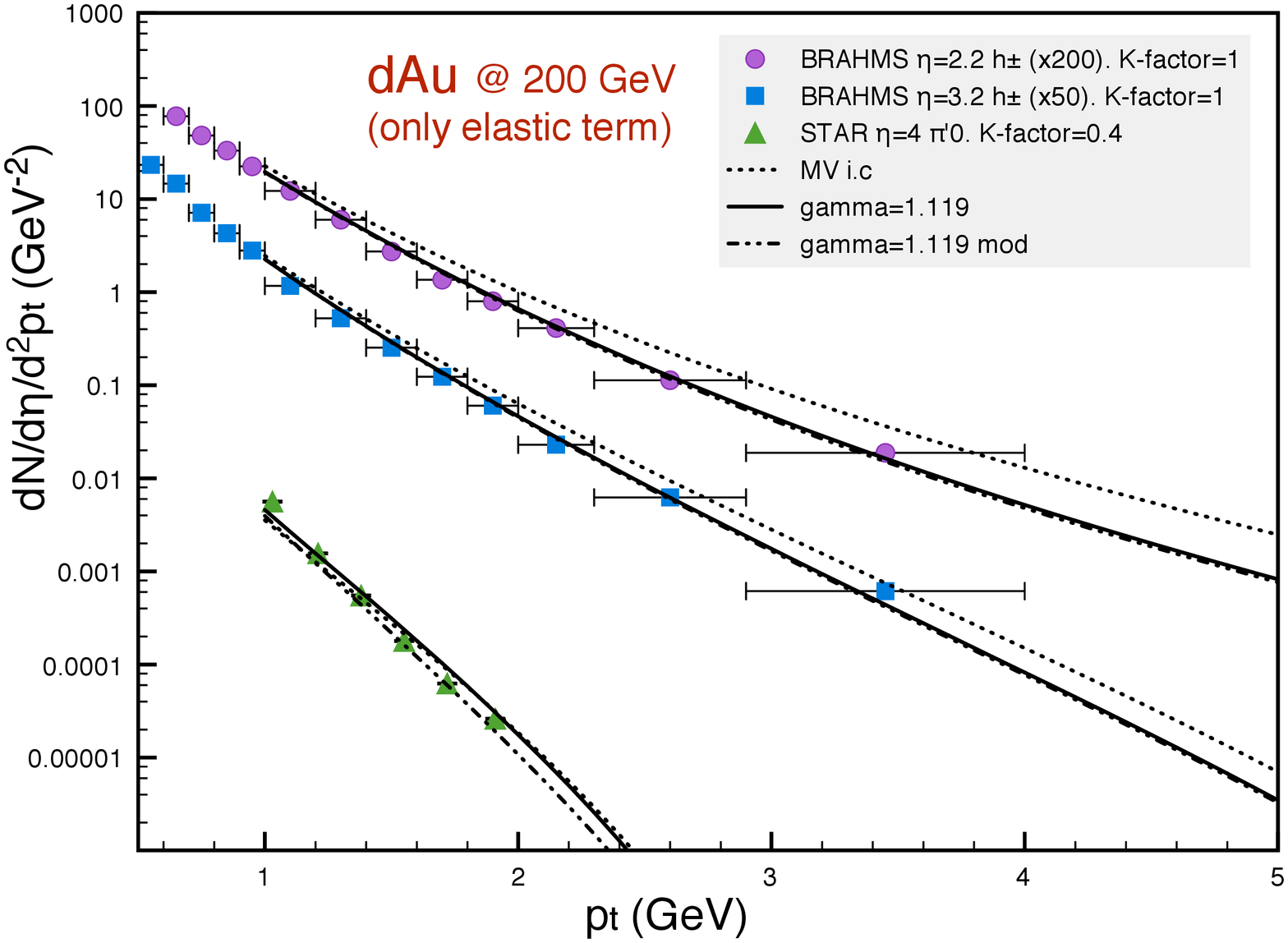}
\end{center}
\vspace*{-0.4cm}
\caption[a]{Comparison of the rcBK-MC results obtained with only the
  elastic term of the hybrid formalism, Eq.~(\ref{hybel}), to the RHIC
  forward data on single inclusive charged hadron (BRAHMS
  data~\cite{Arsene:2004ux}) and neutral pion yields in p+p (left) and
  d+Au collisions (right). Solid lines correspond to the
  $\gamma=1.119$ i.c., dashed-dotted to also $\gamma=1.119$ i.c but using the prescription in Eq. (\ref{eq:addQs2iniAAMQS}) for the initial saturation scale. Dotted lines correspond to MV i.c.}
\label{fig:RHICfwd-el}
\end{figure}
In Fig.~\ref{fig:RHICfwd-el} we compare our results for single
inclusive charged hadron (BRAHMS data~\cite{Arsene:2004ux}) and
neutral pion (STAR data~\cite{Adams:2006uz}) distributions measured in
p+p and d+Au collisions at RHIC. In this figure we include only the
elastic component of the hybrid formalism. In what follows we adopt
the DSS-NLO fragmentation functions as the default ones for all the
calculations performed within the hybrid formalism. Our results show a
good agreement with data. However, the figure also illustrates that
RHIC forward data does not constrain well the initial conditions for
the evolution of nuclear wave functions: both the UGD MV and g1119
sets (using either the {\it natural}, Eq.\ (\ref{eq:addQs2ini}), or
the {\it modified}, Eq.\ (\ref{eq:addQs2iniAAMQS}), ansatz for the
initial saturation scale at every point in the transverse plane) yield
a comparably good description of data. This is due to the fact that
transverse momentum distributions in the forward region do not probe
the $k_T\gg Q_s$ tails of the UGDs.

Similar to previous phenomenological works, we found that no K-factors
are needed to describe data at rapidities $\eta=2.2$ and 3. However,
STAR data at more forward rapidities can only be well described if a
K-factor $\approx0.4$ is introduced. This may be an indication that
large-$x$ phenomena non included in the CGC may be relevant in the
region close to the kinematic limit of phase space. Note, however,
that the value of the $K$-factor depends significantly both on the
UGD and on the FF.

\begin{figure}[htb]
\begin{center}
\includegraphics[width=0.49\textwidth]{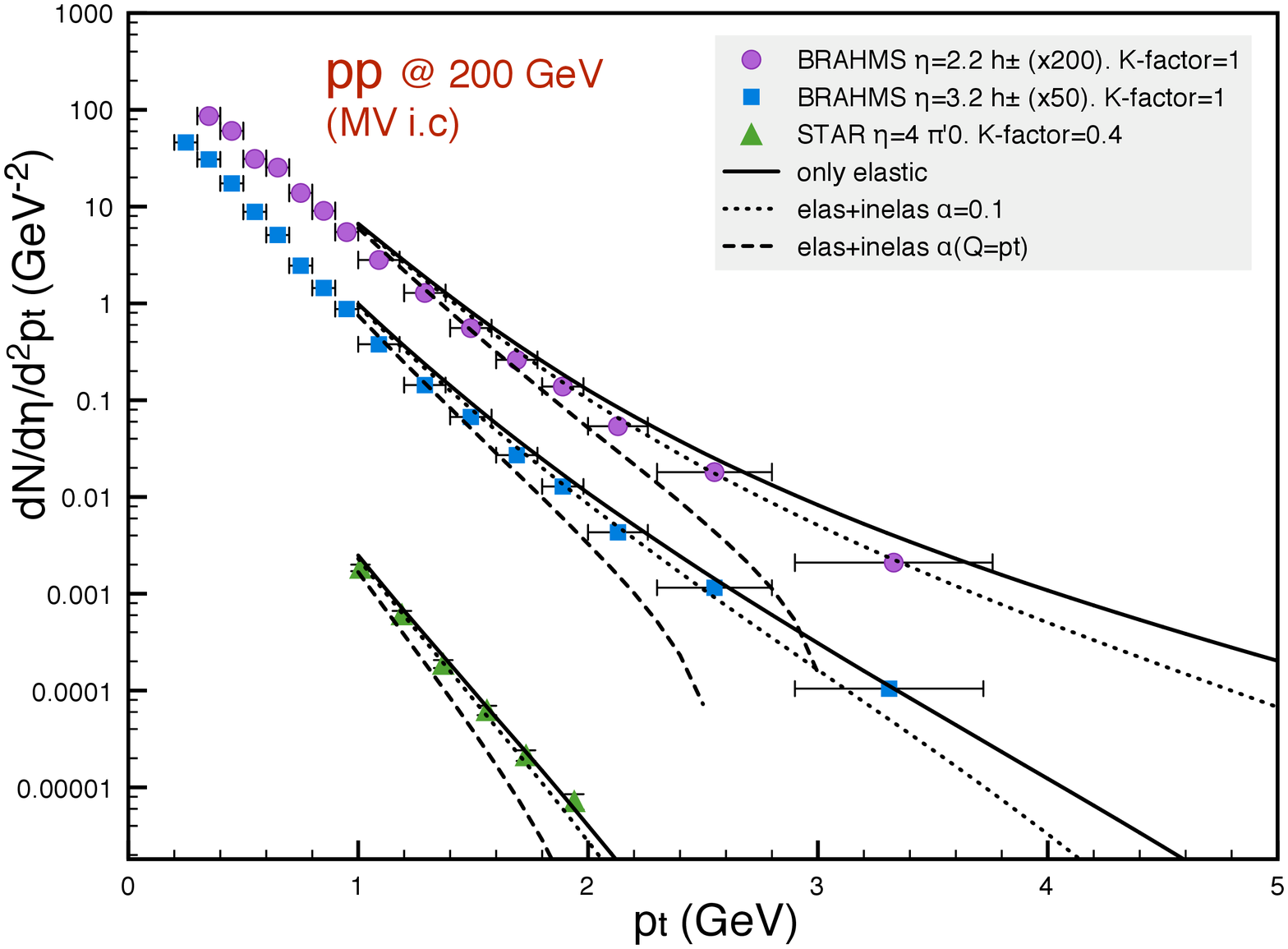}
\includegraphics[width=0.49\textwidth]{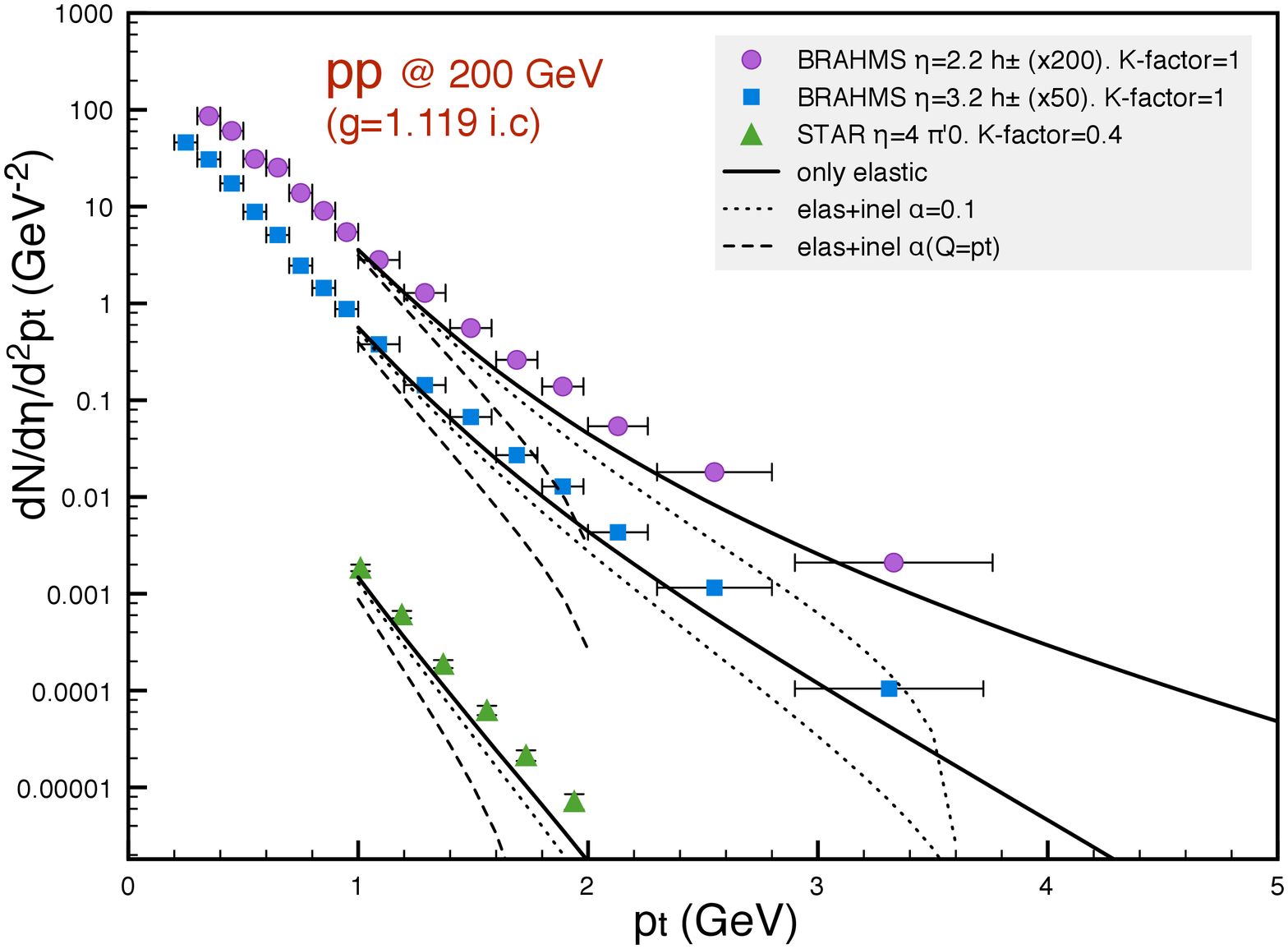}
\includegraphics[width=0.49\textwidth]{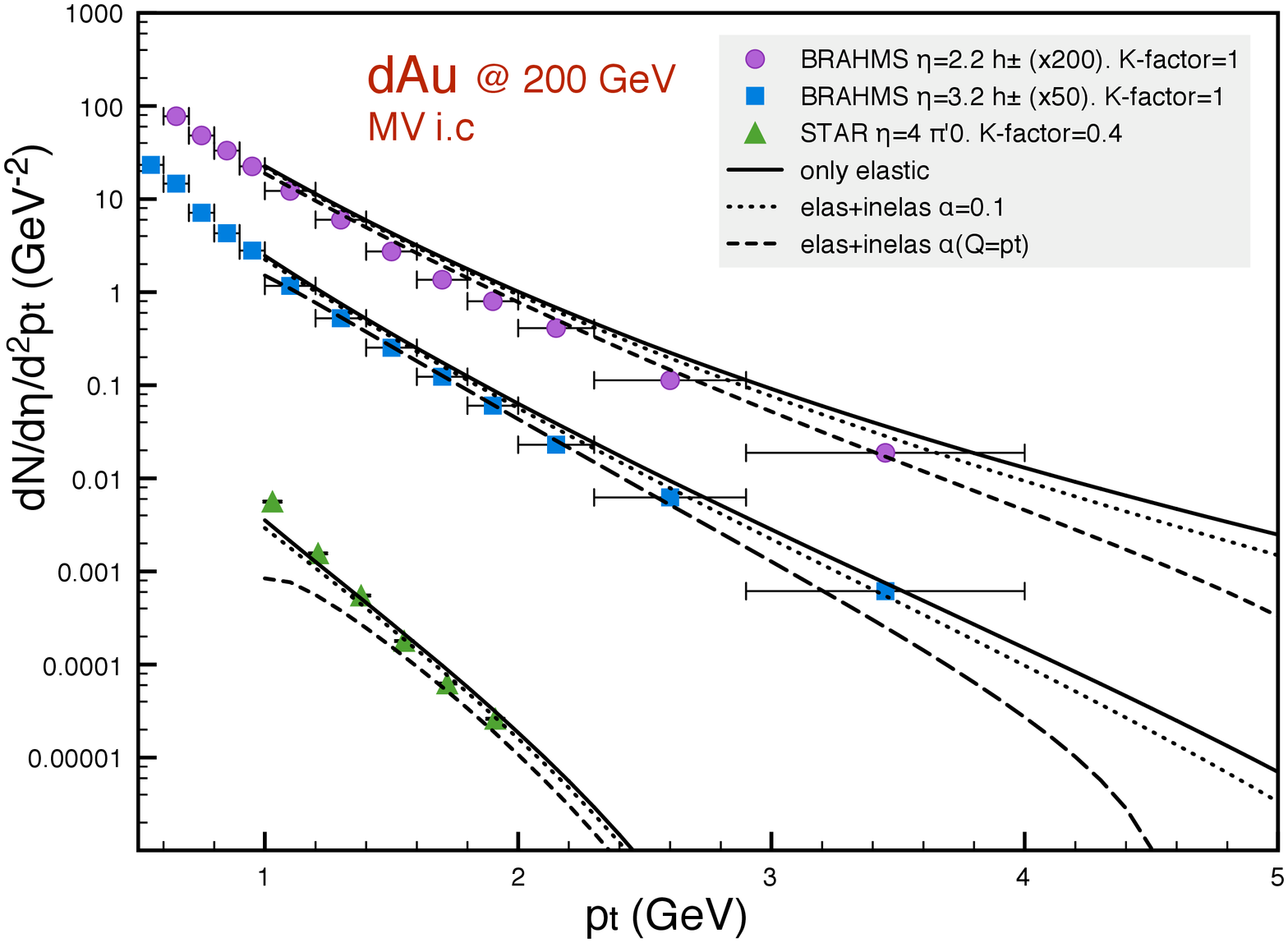}
\includegraphics[width=0.49\textwidth]{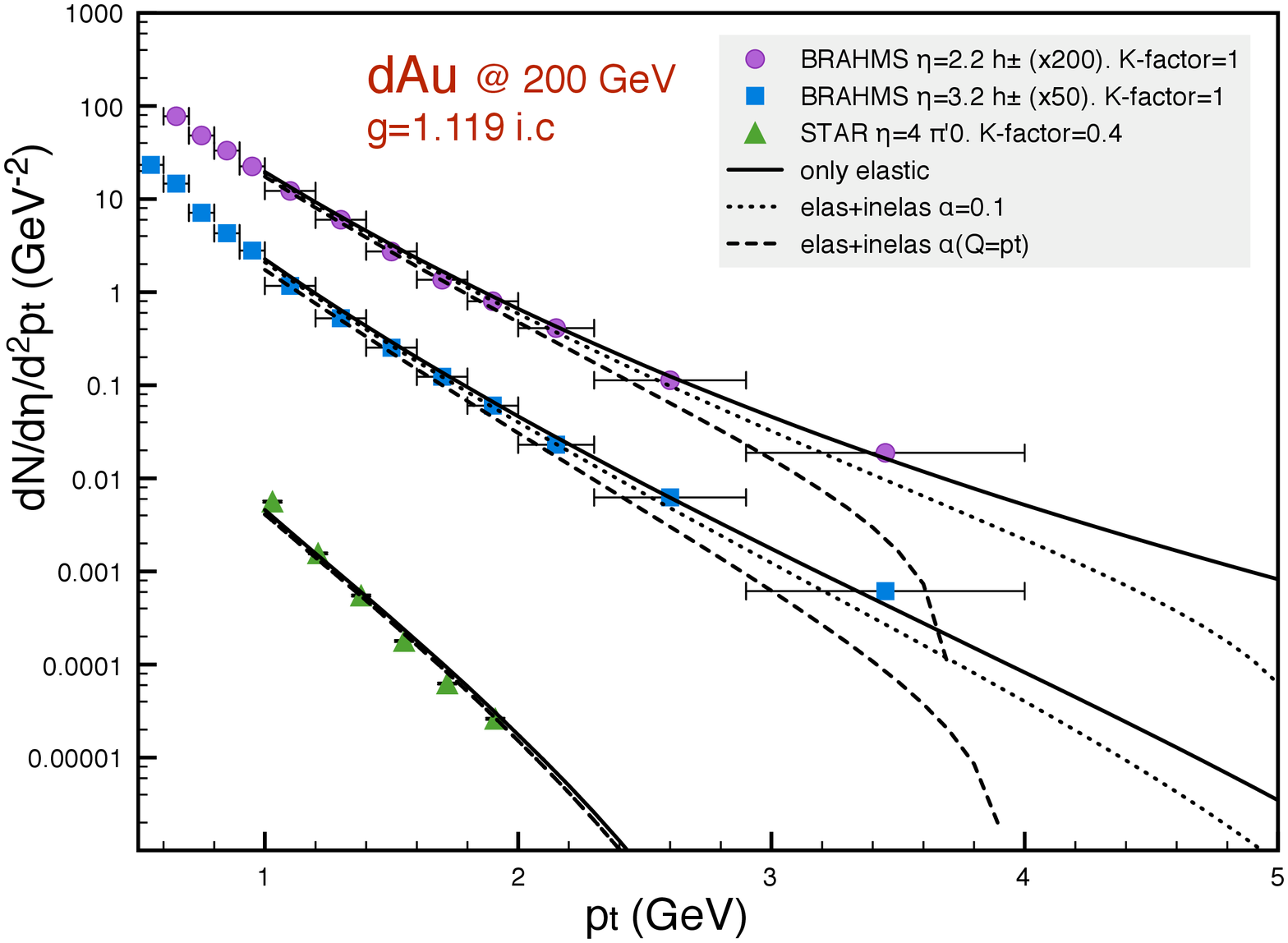}
\end{center}
\vspace*{-0.4cm}
\caption[a]{Same as Fig.~\ref{fig:RHICfwd-el} but including the
  inelastic term in the hybrid formalism. Solid lines are the same as
  in Fig.\ \ref{fig:RHICfwd-el}. Dotted and dashed lines correspond to
  $\alpha_{s}=0.1$ and $\alpha_{s}=\alpha_{s}(Q=p_{t})$ in
  Eq.\ (\ref{hybinel}), respectively.}
\label{fig:RHICfwd-inel}
\end{figure}

In Fig.\ \ref{fig:RHICfwd-inel} we show the comparison to the same
RHIC forward data, now also including the inelastic term in the hybrid
formalism. We explore both fixed $\alpha_{s}=0.1$ as well as one-loop
running coupling at the scale $Q$. We observe that the effect of this
additional term can be very large, especially at large transverse
momentum. We note that, despite the fact that the coupling decreases
with increasing transverse momentum, the running coupling prescription
causes a larger effect than the fixed coupling one. 

We observe that the inelastic term exhibits a harder $p_T$-dependence
than the elastic contribution, and at some transverse momentum it
overwhelms the elastic contribution.  The crossing point depends on
the particular choice of UGD.  The effects from the inelastic
corrections are stronger for the steeper g1119 initial conditions than
for the MV ones over the entire range of transverse momentum shown in
Fig.~\ref{fig:RHICfwd-inel}.  Also, the importance of the inelastic
term depends on the collision system or, equivalently, on the target
saturation scale: it is stronger for p+p than for d+Au collisions. For
p+p collisions in particular it appears that the present formalism
does not provide a stable result as the inelastic correction
overwhelms the leading elastic contribution already at moderate values
of transverse momentum. This not a completely unexpected result since,
parametrically, the inelastic term is proportional to
$\ln(p_{t}/Q_{st})$, with $Q_{st}$ the target saturation scale, while
the elastic term scales as $\ln(p_{t}/\Lambda_{QCD})$ (see discussion
in~\cite{Altinoluk:2011qy}).  Given the importance and magnitude of
the inelastic term, our findings call for a complete phenomenological
analysis of the full NLO corrections.

\begin{figure}[htb]
\begin{center}
\includegraphics[width=0.48\textwidth]{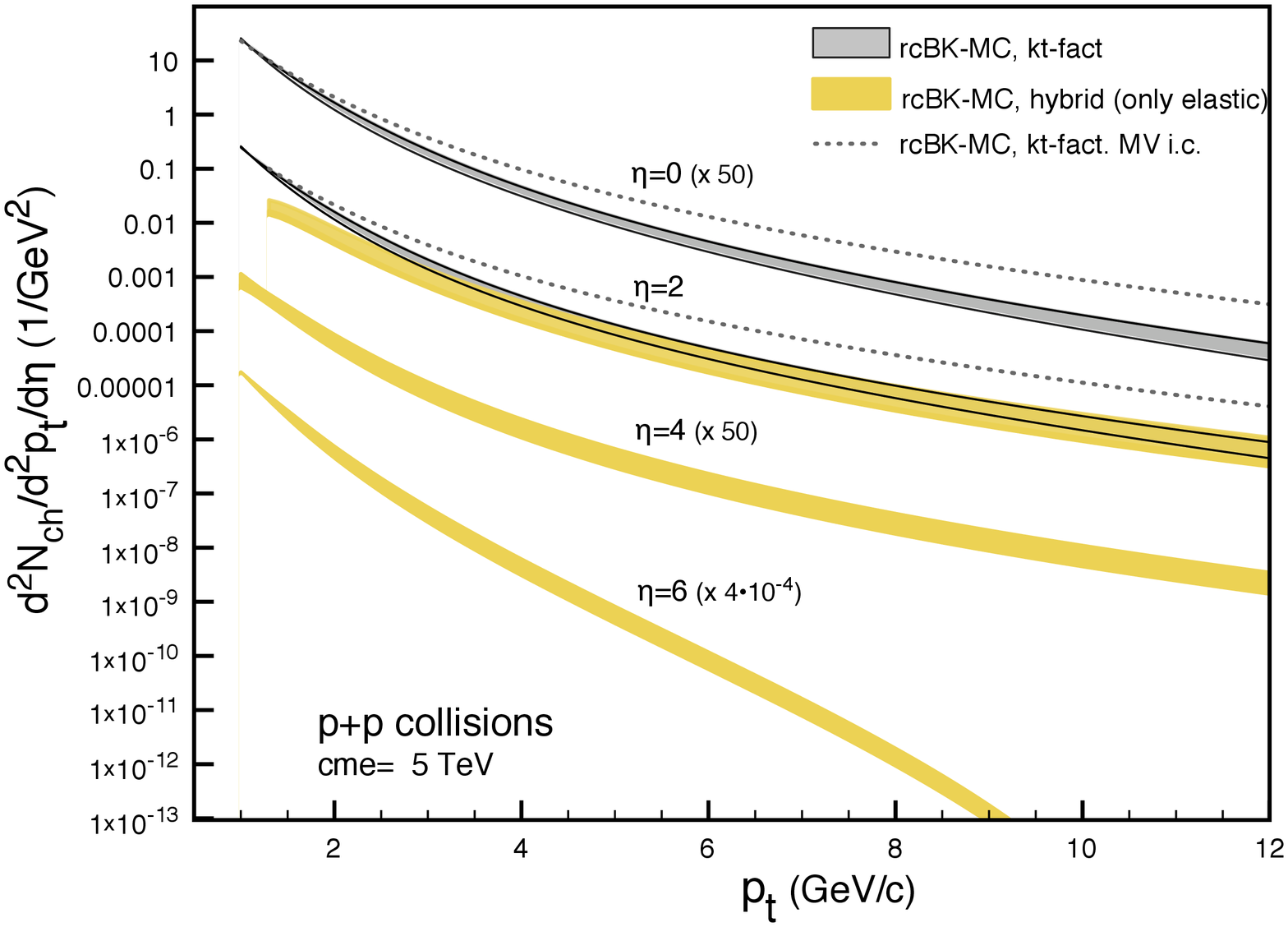}
\includegraphics[width=0.48\textwidth]{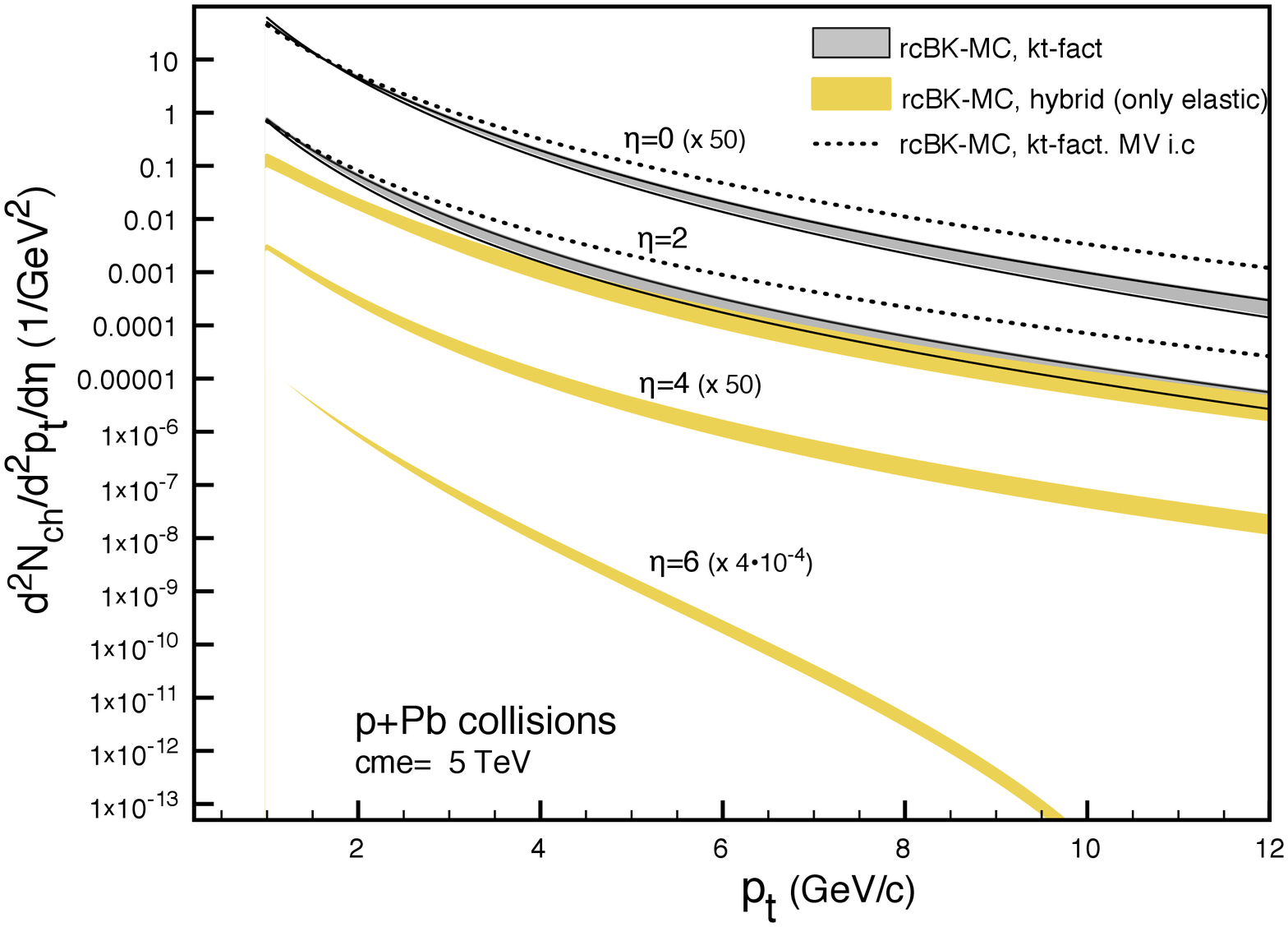}
\end{center}
\vspace*{-0.4cm}
\caption[a]{Predictions for the single
  inclusive charged hadrons yields in p+p and minimum-bias p+Pb collisions at 5~TeV
  collision energy at rapidities 0, 2, 4 and 6. The grey bands at y=0
  and 2 correspond to the rcBK-MC results using kt-factorization,
  Eq.\ (\ref{kt2}). Dotted lines correspond to MV i.c.. In turn, the yellow bands at $\eta=2$, 4 and 6 have been
  obtained using the LO hybrid formalism including only the {\it elastic} term, Eq.\ (\ref{hybel}). The overall normalization has been adjusted by a factor 50, 1, 0.02 and $4\cdot 10^{-4}$ for rapidities 0, 2, 4 and 6 respectively to improve the visibility. } 
  \label{fig:spectraLHC}
\end{figure}

\begin{figure}[htb]
\begin{center}
\includegraphics[width=0.48\textwidth]{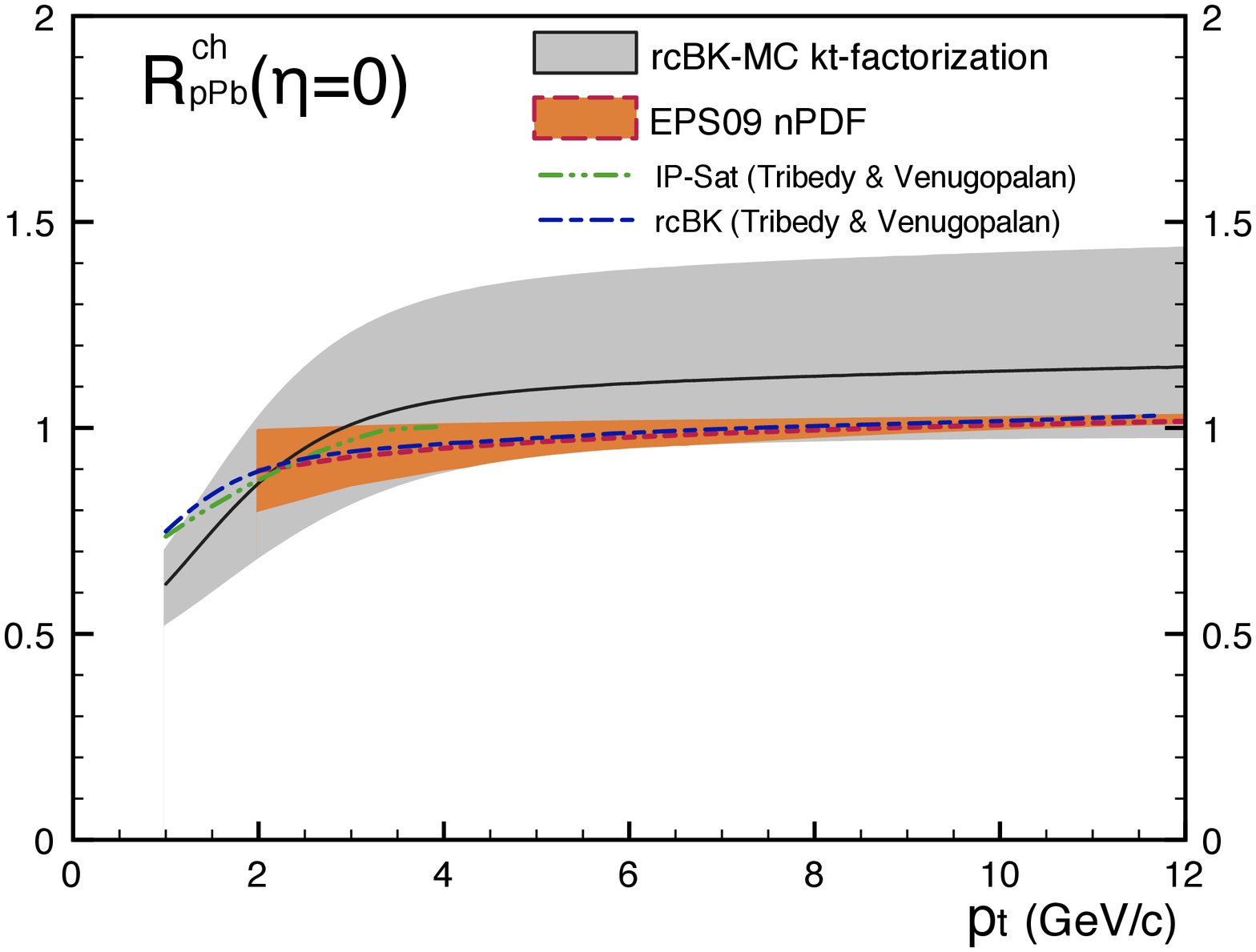}
\includegraphics[width=0.48\textwidth]{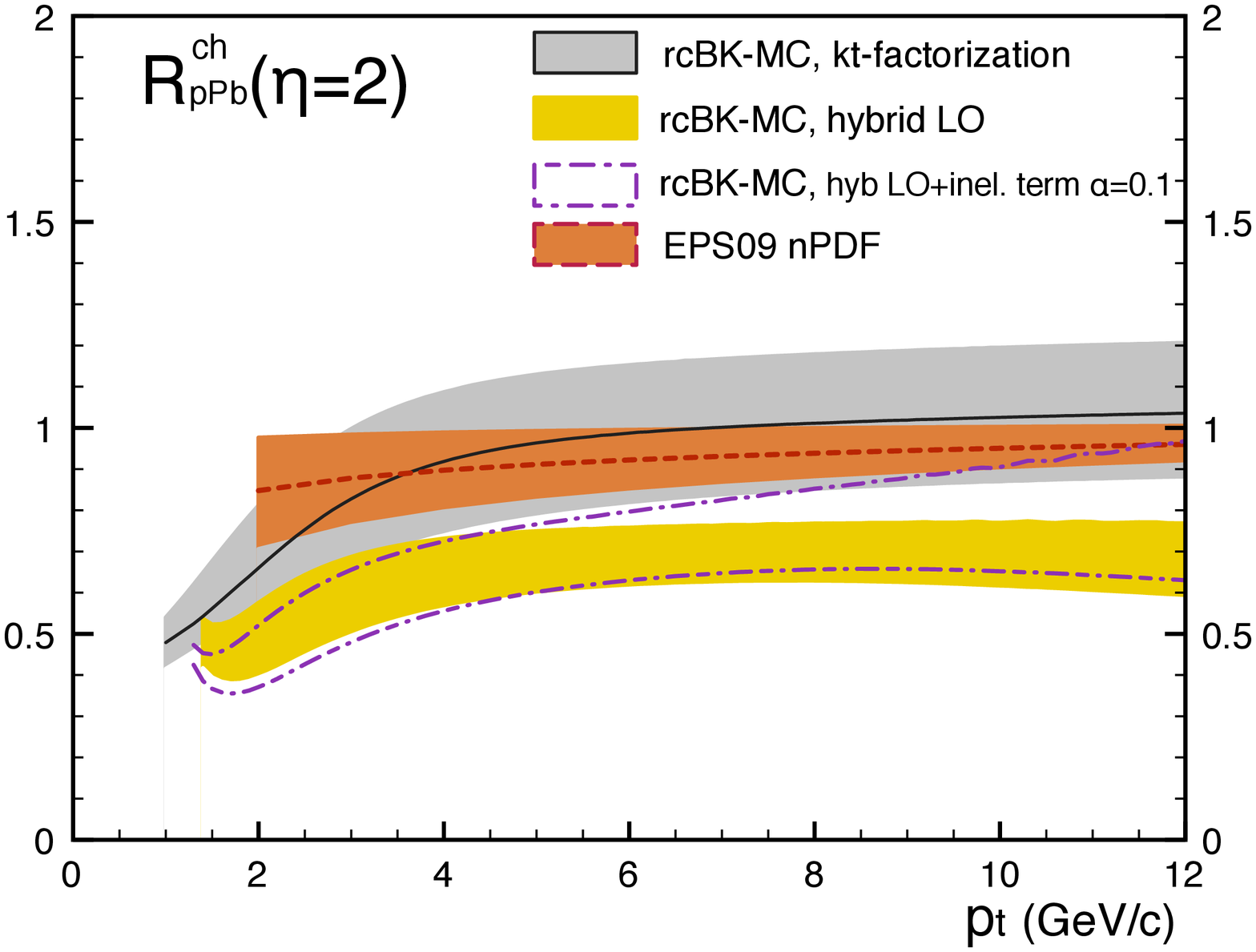}
\includegraphics[width=0.48\textwidth]{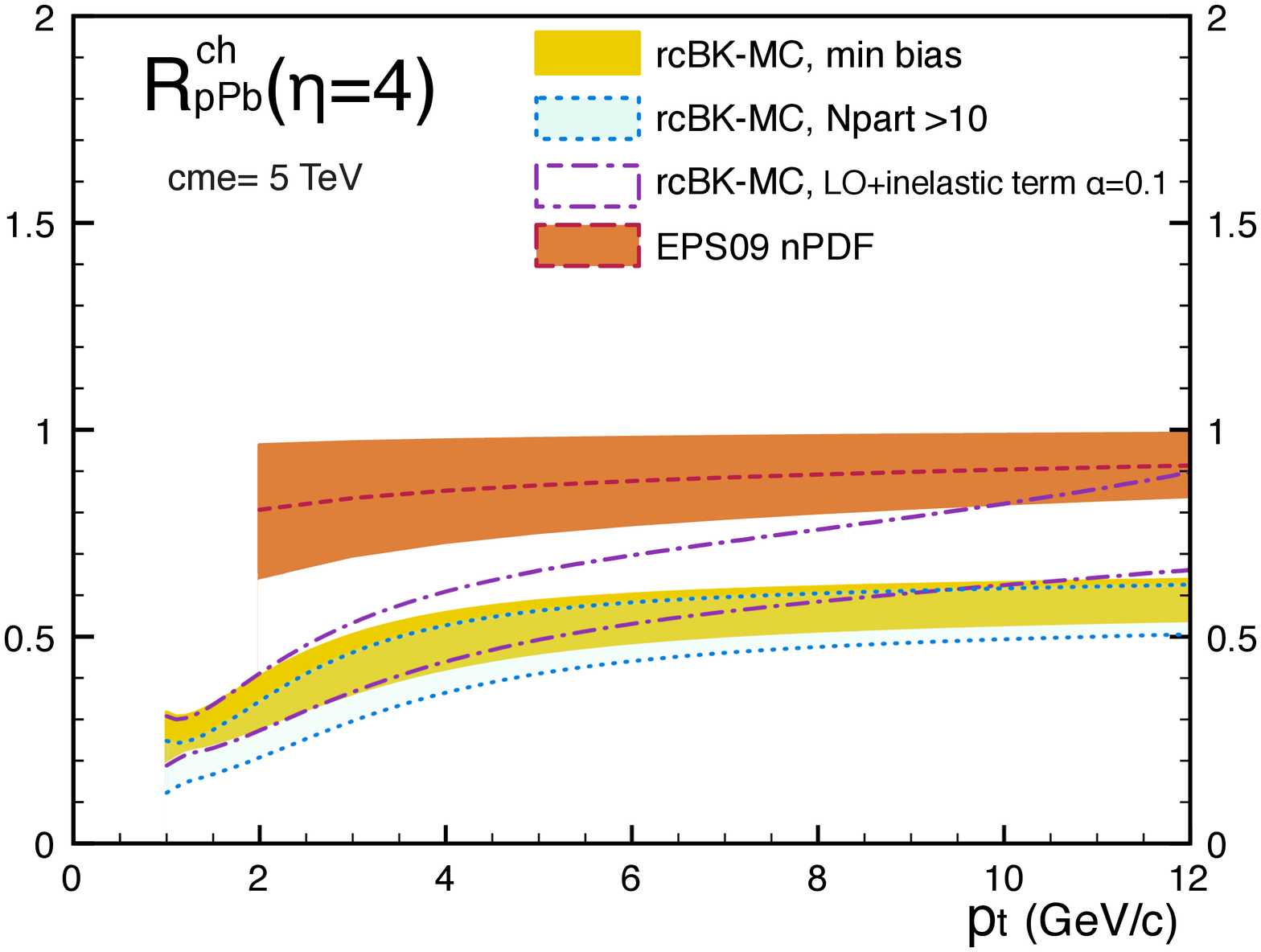}
\includegraphics[width=0.48\textwidth]{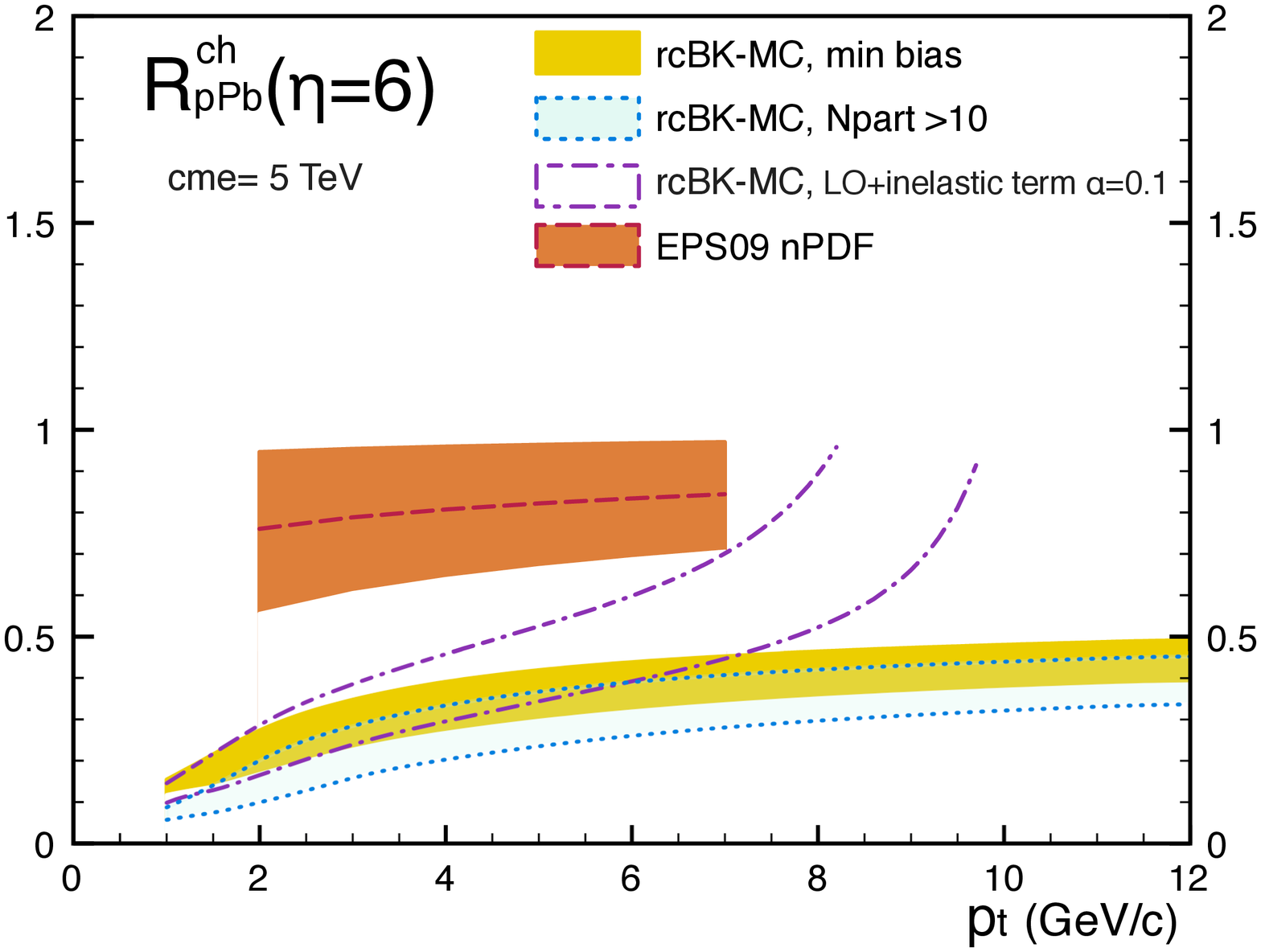}
\end{center}
\vspace*{-0.4cm}
\caption[a]{The nuclear modification factor $R_{\rm p+Pb}$ for single
  inclusive charged hadrons in minimum-bias p+Pb collisions at 5~TeV
  collision energy at rapidities 0, 2, 4 and 6. The grey bands at y=0
  and 2 correspond to the rcBK-MC results using kt-factorization,
  Eq.\ (\ref{kt2}). In turn, the yellow bands at $\eta=2$, 4 and 6 have been
  obtained using the LO hybrid formalism, Eq.\ (\ref{hybel}), in
  minimum bias collisions. The blue bands between the dotted lines also
  correspond to LO hybrid results for collisions with a
  centrality cut $N_{\rm part}>10$. Finally the dashed dotted curves at
  $\eta=2$, 4 and 6 correspond to minimum bias collisions calculated within
  the hybrid formalism incl.\ the inelastic term from
  Eq.~(\ref{hybinel}) with $\alpha_{s}=0.1$. }
\label{fig:RpPbmb}
\end{figure}
We now proceed to p+Pb collisions at LHC energy, $\surd s=5$~TeV.
In Figs.~\ref{fig:spectraLHC} and \ref{fig:RpPbmb} we show our results for the single
inclusive charged hadrons yields in p+p and minimum bias p+Pb collisions and the nuclear
modification factor $R_{\rm p+Pb}$ for minimum bias collisions respectively. We
compare also to $R_{\rm p+Pb}$ from collinear factorization using
EPS09 nPDFs~\cite{Eskola:2009uj,QuirogaArias:2010wh} as well as to
results from the ``IP-sat'' model and from an independent rcBK
implementation\footnote{To mention two differences to our work:
  ref.~\cite{Tribedy:2011aa} uses a different fragmentation function
  and does not treat fluctuations of the nucleon configurations in the
  target. The predictions are not far apart but the difference
  illustrates the sensitivity of $R_{\rm pA}$ to such
  ``details''.}~\cite{Tribedy:2011aa}.

\begin{figure}[htb]
\begin{center}
\includegraphics[width=0.48\textwidth]{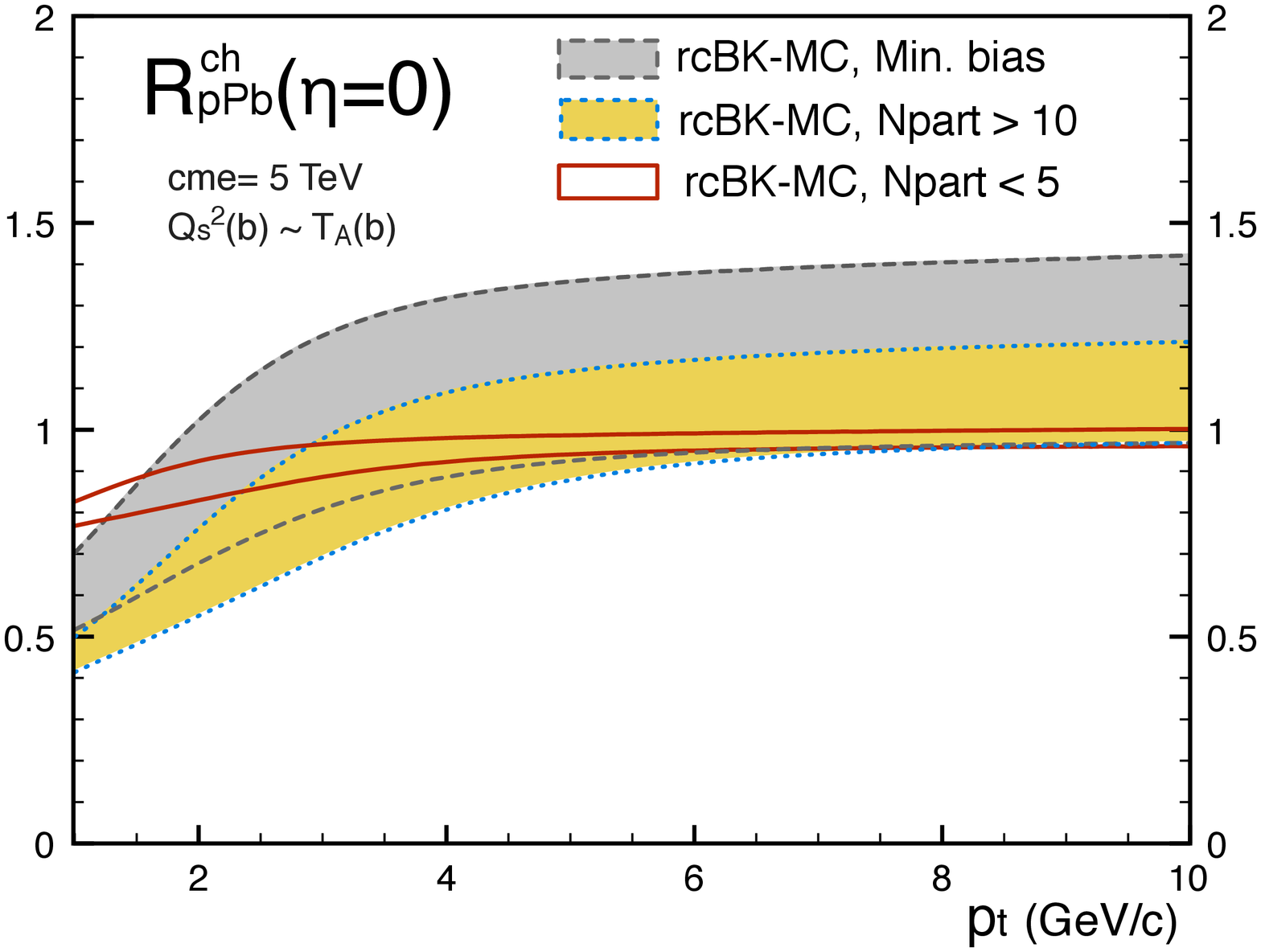}
\includegraphics[width=0.48\textwidth]{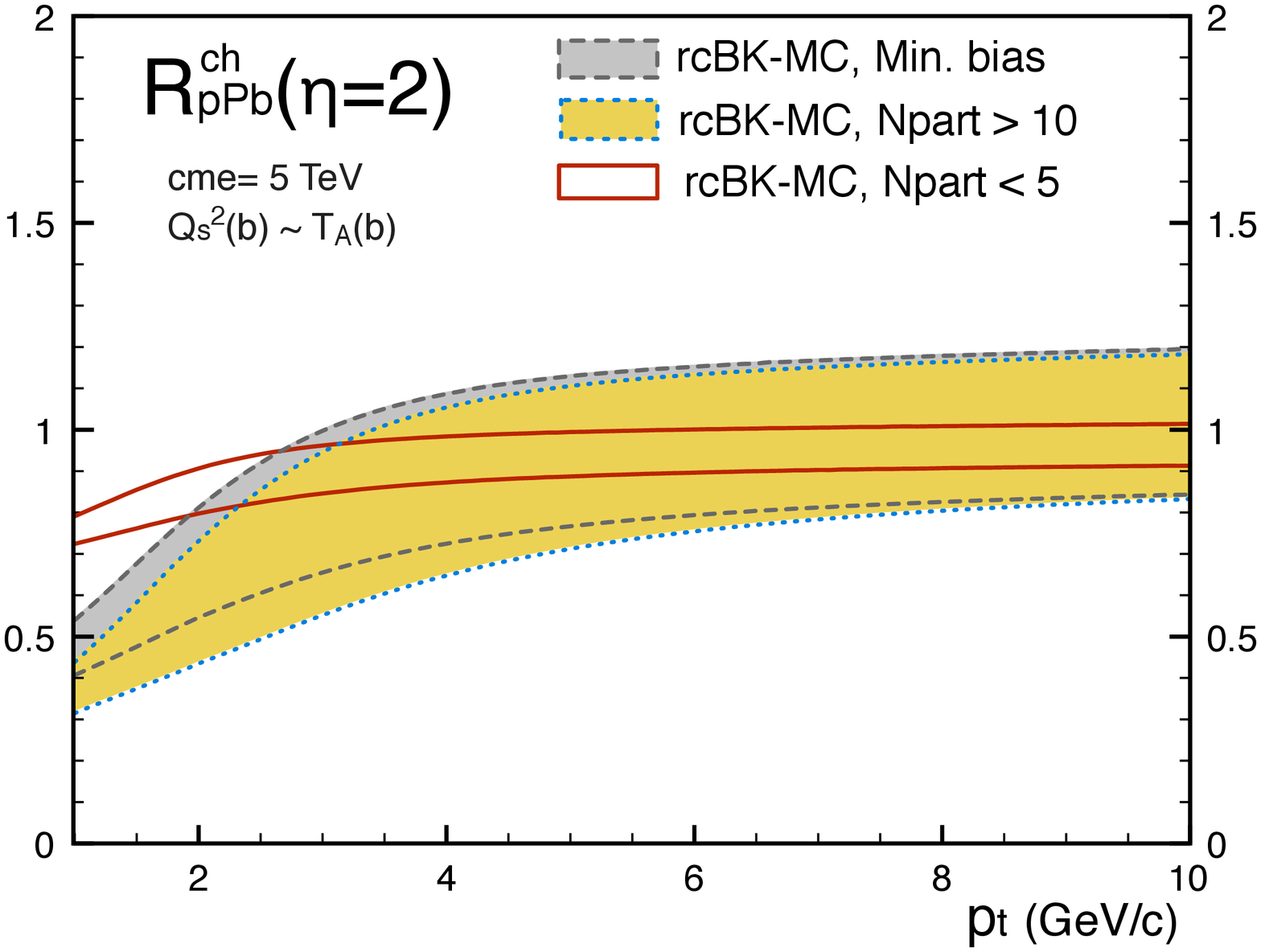}
\end{center}
\vspace*{-0.4cm}
\caption[a]{Same as Fig.~\ref{fig:RpPbmb} for two different centrality
  classes. Here, for all curves the initial saturation momentum
  squared in the Pb target is taken to be proportional to the density
  of nucleons per unit transverse area in a given event,
  $Q_s^2(x_0;{\bf b})\sim T_A({\bf b})$, according to the {\it
    natural} prescription given by Eq.~(\ref{eq:addQs2ini}). Results
  in this plot have been calculated in the $k_{t}$-factorization
  formalism.}
\label{fig:RpPb_centr}
\end{figure}

\begin{figure}[htb]
\begin{center}
\includegraphics[width=0.48\textwidth]{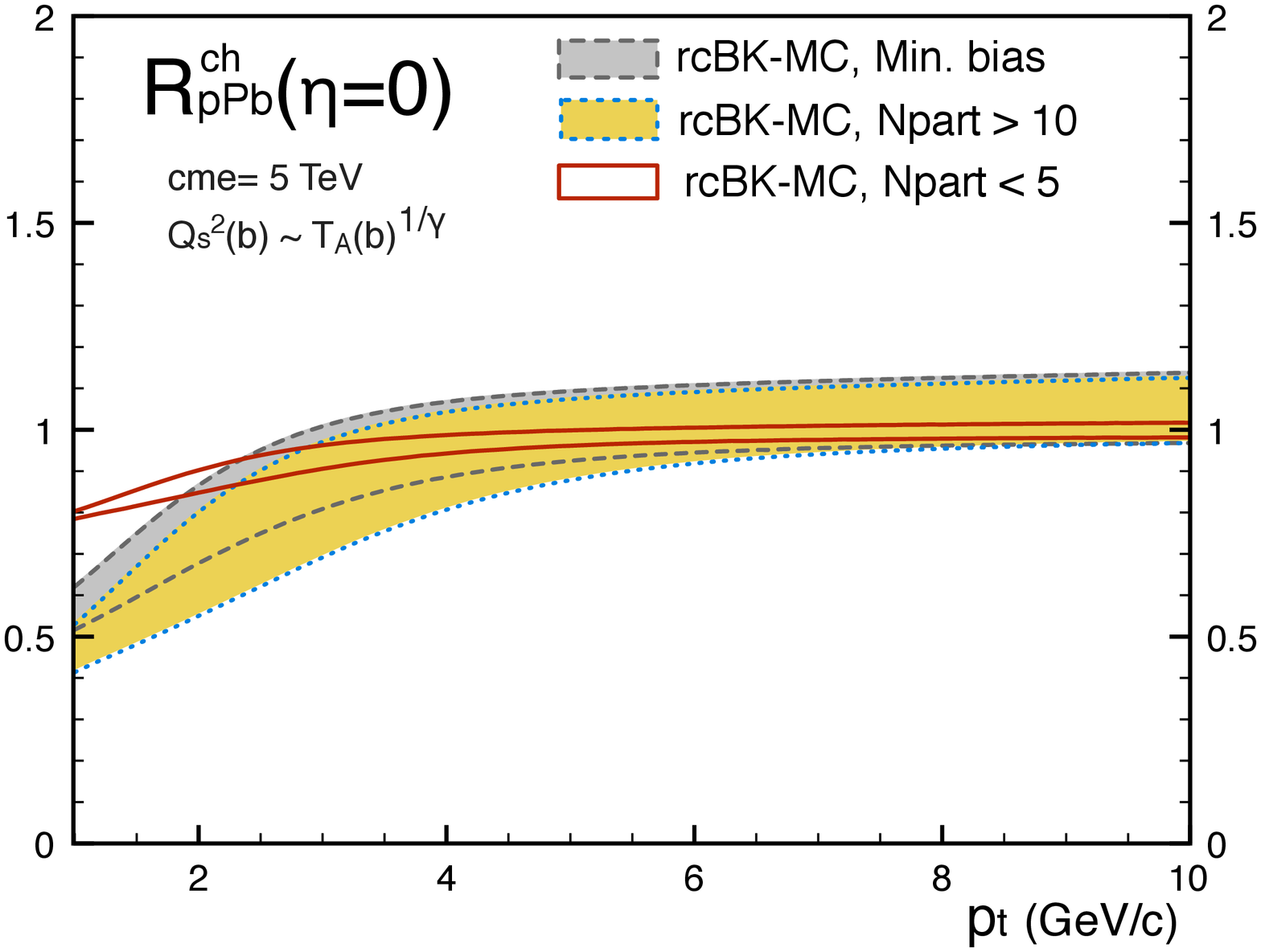}
\includegraphics[width=0.48\textwidth]{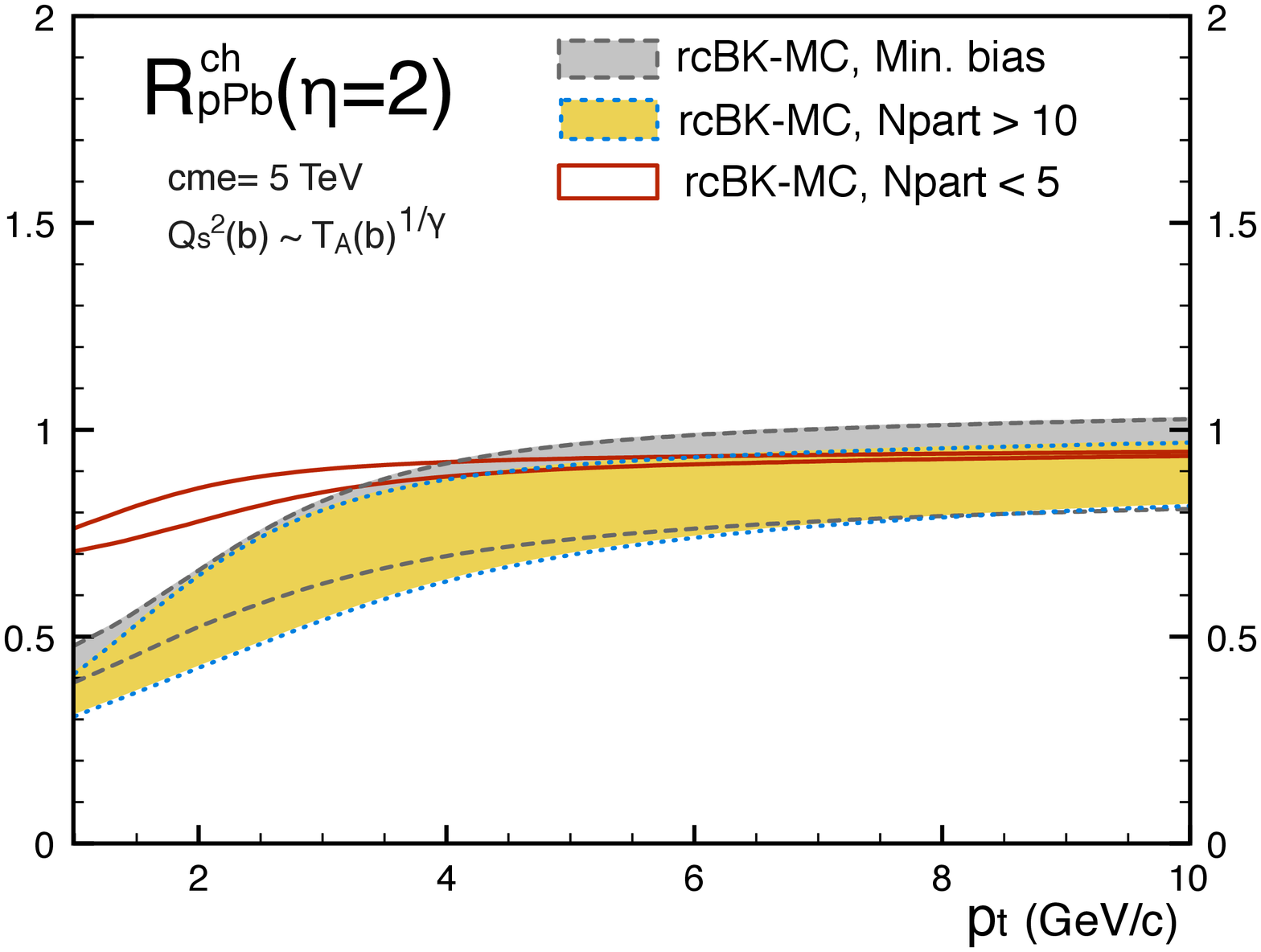}
\end{center}
\vspace*{-0.6cm}
\caption[a]{Same as Fig.~\ref{fig:RpPb_centr} but here the initial
  saturation momentum squared in the Pb target is taken as
  $Q_s^2(x_0;{\bf b})\sim T_A^{1/\gamma}({\bf b})$, according to the
  {\it modified} prescription given by
  Eq.~(\ref{eq:addQs2iniAAMQS}). Results in this plot have been
  calculated in the $k_{t}$-factorization formalism.}
\label{fig:RpPb_centrGamma}
\end{figure}

Before discussing the results let us first explain the meaning of the
rcBK-MC bands shown in
Figs.~\ref{fig:RpPbmb}-\ref{fig:RpPb_centrGamma}: They comprise the
results for $R_{\rm pPb}$ calculated according to Eq.~(\ref{eq:RpPb})
using the three UGD sets (g1119, g1101 and MV), the three kind of
fragmentation functions (KKP-LO, DSS-LO and DSS-NLO) and the two
possibilities to determine the initial saturation scale ({\it
  natural}, Eq.~(\ref{eq:addQs2ini}), or {\it modified},
Eq.~(\ref{eq:addQs2iniAAMQS})) considered throughout this work, always
using the same configuration in the numerator --p+Pb-spectrum-- and
denominator --p+p-spectrum--. The upper limit of the bands correspond
in all cases to $R_{\rm pPb}$ calculated with UGD set g1.119 together
with the {\it natural} prescription for the initial saturation
scale. The black solid line in the plots for $\eta=0$ and 2 in
Fig.~\ref{fig:RpPbmb} represents the upper limit of the band if only
{\it modified} initial conditions are used (such distinction is not
necessary for Figs.~\ref{fig:RpPb_centr} and \ref{fig:RpPb_centrGamma}
since both cases are treated separately). For the results obtained
within the $k_{t}$-factorization formalism the upper limit correspond
to KKP-LO fragmentation functions, while for the results obtained
within the hybrid formalism (both for {\it only elastic} and {\it
  elastic +inelastic} curves) the upper limit of the bands corresponds
to DSS-NLO fragmentation functions. In turn, the lower limits of the
bands correspond in all cases to UGD set MV and DSS-NLO fragmentation
functions. The results for all other possible configurations --
i.e.\ other UGDs and fragmentation functions and choice of {\it
  natural} or {\it modified} initial conditions-- fall within the
plotted bands; individual curves are not shown for clarity of the
presentation.

We observe a suppression of $R_{\rm p+Pb}=0.6\pm0.1$ at midrapidity
and $p_t=1$~GeV. Over the entire range of $p_t$ shown in the figure,
and for both UGDs, $R_{\rm p+Pb}$ decreases with increasing rapidity
(towards the proton fragmentation region).  This is consistent with
expectations from non-linear evolution which suppresses additional
gluon emissions in dense wave functions as compared to the dilute
limit.  However, we find that $R_{\rm p+Pb}$ increases with $p_t$ and
reaches, or even exceeds, unity for hadron transverse momenta of
several GeV.

This is due to the following effects. First of all, the light-cone
momentum fractions of gluons which contribute to hadron production at
fixed rapidity increase proportional to $p_t$. This drives us closer
to the initial condition which in fact exhibits an ``anti-shadowing''
{\it enhancement} of the UGD tail at high intrinsic transverse
momentum\footnote{Note, however, that the UGD with $\gamma=1$ MV-model
  initial condition is very close to the EPS09 nPDF at high transverse
  momentum and $y=0$. This indicates that for this UGD
  ``anti-shadowing'' is weak.}. This higher-twist ``anti-shadowing''
component is further enhanced by Glauber fluctuations in the Pb target
since $\varphi(x,k_\perp)$ is not linear in the target thickness while
$\langle N_{\rm coll}\rangle$ in the denominator of
Eq.~(\ref{eq:RpPb}) is. Nevertheless, it is remarkable that small-$x$
quantum evolution in the rcBK approximation predicts a disappearance
at LHC energies (and $\eta\ge0$) of the clear Cronin peak observed so
far in all proton-nucleus collisions at lower energies\footnote{Such a
Cronin peak at 5~TeV is also predicted by some leading-twist shadowing
models including parton intrinsic transverse momenta~\cite{Xu:2012vw}.}. 
Our results confirm earlier more schematic or qualitative predictions of
this effect~\cite{Albacete:2003iq,Kharzeev:2003wz}. Forthcoming LHC
data will therefore provide a very important test for the evolution
speed predicted by the running coupling BK equation.

It is also remarkable that at intermediate rapidities $\eta\approx2$
the hybrid formalism with only the elastic term gives $R_{\rm pPb}$
systematically below the one from $k_{t}$-factorization.  However, the
inclusion of the inelastic term Eq.\ (\ref{hybinel}) in the
calculations within the hybrid formalism tends to systematically
increase $R_{\rm pPb}$ as compared to only the elastic component,
bringing it closer to the $R_{pPb}$ obtained from
$k_{t}$-factorization.  This effect is visible, in particular, at high
transverse momentum, lifting $R_{pPb}$ closer to unity. This confirms
the expectation that higher-order and energy conservation corrections
should bring $p_{t}$-spectra closer to the DGLAP limit at high
transverse momenta, where gluon densities are smaller and non-linear
corrections should be less relevant. However, as discussed previously,
the precise quantitative effect of the inelastic term appears to
depend quite strongly on the value of the strong coupling, preventing
us to make precise quantitative predictions (only the results
corresponding to $\alpha_{s}=0.1$ are shown in
Fig.\ (\ref{fig:RpPbmb}); those corresponding to
$\alpha_{s}=\alpha_{s}(Q)$ produce a much larger effect on $R_{\rm
  pPb}$). Also, it is somewhat surprising that the effect is strongest
at the most forward rapidities $\eta\approx6$ and the highest
transverse momentum (we cut off the curves corresponding to the
inelastic term at $p_{t}\approx 8$ in this plot, since they grow very
fast above unity for larger $p_{t}$). However, this region
($\eta\gtrsim6$, $p_{t}\gtrsim 8$ GeV) is close to the kinematic
limit at LHC energies (see Fig.\ \ref{fig:x2Cover}). It could be
that, similar to our discussion of forward RHIC data, the CGC
formulation discussed in this work is not complete or inaccurate
around the limit of phase space, hence the uncontrolled growth of
$R_{\rm pPb}$ at the most forward rapidities.

In Fig.~\ref{fig:RpPb_centr} we show $R_{\rm p+Pb}$ for two different
centrality classes selected according to the number of participant
nucleons\footnote{In p+A collisions it is not straightforward
  experimentally to perform centrality selection via impact parameter
  cuts. Also, because of large fluctuations impact parameter bins
  correspond to rather broad distributions of $N_{\rm part}$.}. At
$p_t=1$~GeV we observe the expected pattern of
stronger suppression (smaller $R_{\rm p+Pb}$) for more central
collisions. In the $N_{\rm part}>10$ centrality class suppression now
persists up to $p_t= 2-3$~GeV. 

For the UGD with $\gamma=1$ MV-model initial condition (lower end of
the bands in Fig.~\ref{fig:RpPb_centr}) one observes, generically, the
expected pattern: i) at $y=0$ there is suppression at low $p_t$ while
$R_{\rm p+Pb}\to1$ with increasing $p_t$ as the rapidity evolution
window shrinks; ii) there is slightly stronger suppression at low
$p_t$ for $N_{\rm part}>10$ central collisions while the centrality
cut has very little effect at high $p_t$; iii) the suppression
increases with rapidity and $R_{\rm p+Pb}<1$ for all $p_t\lsim10$~GeV
at $y=2$.

The behavior of $R_{\rm p+Pb}$ with AAMQS UGDs ($\gamma=1.119$ initial
condition, upper end of the bands in Fig.~\ref{fig:RpPb_centr}) in
central collisions is more intricate. At $p_t=1$~GeV we still find the
expected decrease of $R_{\rm p+Pb}$ both with centrality and
rapidity. However, for $p_t\gsim4$~GeV we find that $R_{\rm p+Pb}$ is
very similar at $y=0$ and $y=2$.  This UGD exhibits rather
non-linear (in the valence charge density) anti-shadowing at high
intrinsic $k_t$ and so particle production at high $p_t$ in p+Pb
collisions is dominated by fluctuations corresponding to a high
valence charge density in the Pb target (high $N_{\rm part}$). This
can be seen from the fact that at $y=2$ and high $p_t$ there is little
difference between the minimum bias and $N_{\rm part}>10$ centrality
classes.

In Fig.~\ref{fig:RpPb_centrGamma}, finally, we show the nuclear
modification factor for the UGDs with initial saturation momentum
squared $Q_s^2(x_0;{\bf b})\sim T_A^{1/\gamma}({\bf b})$ which
restores the ${\cal N}(r;x_0)\sim r^2$ behavior of the dipole
scattering amplitude in the perturbative limit, $r\, Q_s(x_0)\ll
1$. This prescription for the initial saturation momentum reduces
anti-shadowing at high intrinsic momenta. This leads to slightly lower
values of $R_{\rm p+Pb}$ at high $p_t$ than the AAMQS UGDs
corresponding to $Q_s^2(x_0;{\bf b})\sim T_A({\bf b})$.

\section{Multiplicity and transverse energy in Pb+Pb collisions}

In this section we present the centrality dependence of the
multiplicity and transverse energy in heavy-ion collisions at LHC
energy~\footnote{Very similar results were previously presented
  in~\cite{Albacete:2010ad} (unpublished) before the corresponding
  data was available}. This serves mainly as a rough check for the
dependence of the saturation momentum on the thickness of a
nucleus. We shall find that the present framework leads to a rather
good description of the data. Nevertheless, we recall that we do not
account for final-state effects such as entropy production; also, that
our estimates rely on a crude ``hadronization model'' as well as on
$k_t$-factorization which is not expected to be very accurate for
$p_t$ integrated multiplicities in A+A collisions.  That said, it is
clearly justified to establish that the model is not in gross
contradiction to basic features of heavy-ion data.

\begin{figure}[htb]
\begin{center}
\includegraphics[width=0.48\textwidth]{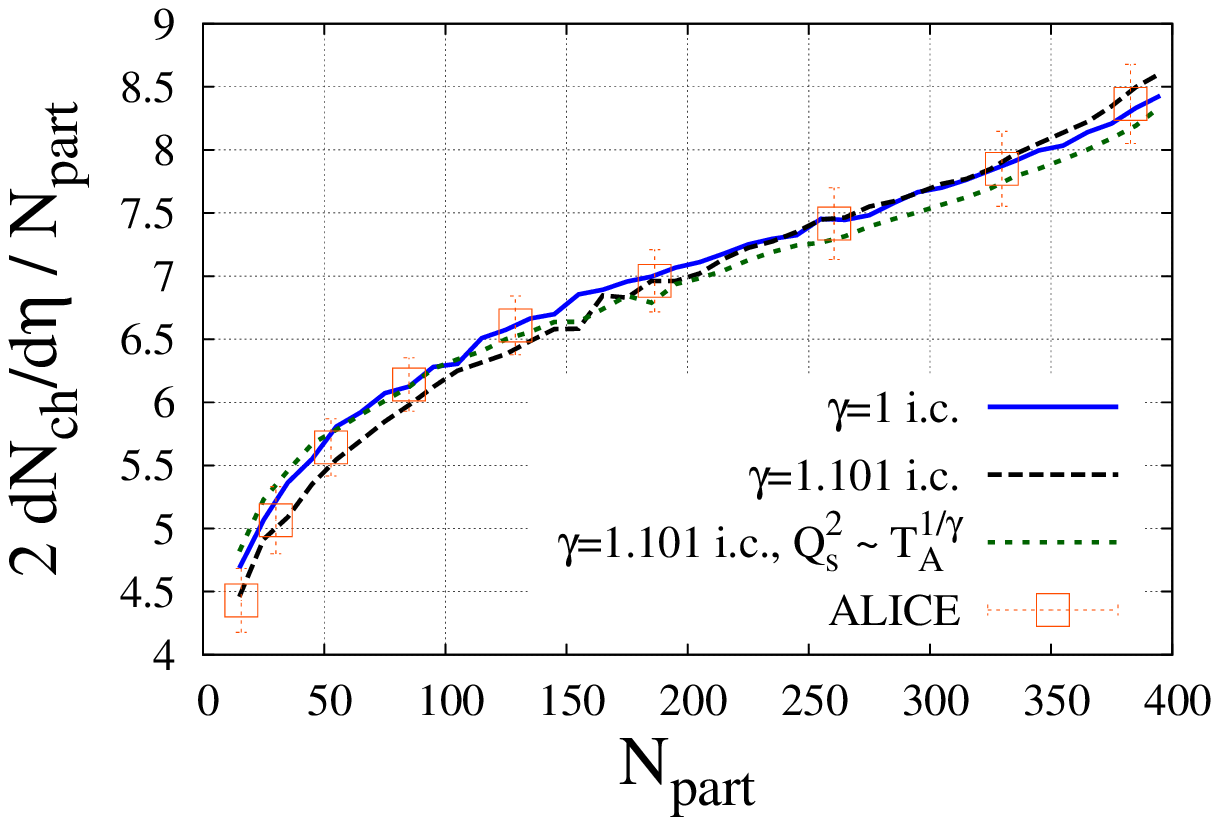}
\includegraphics[width=0.48\textwidth]{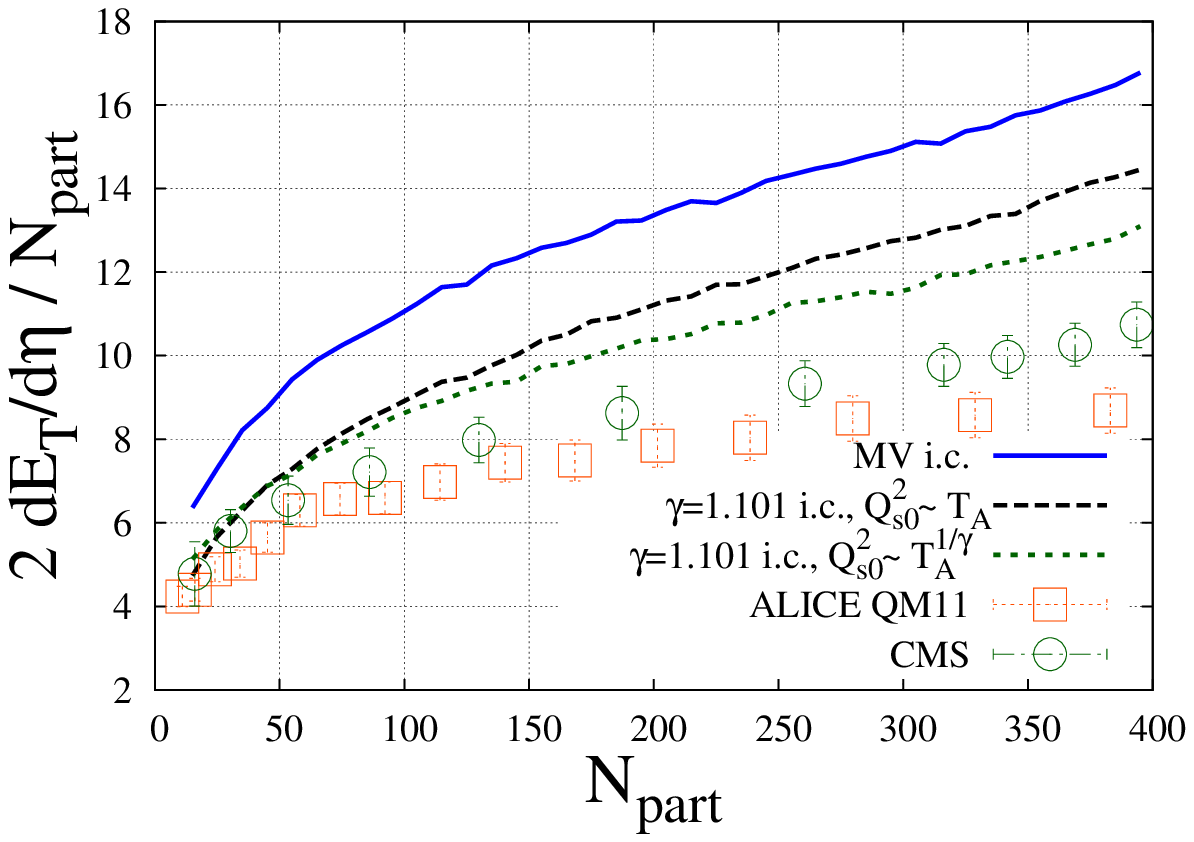}
\end{center}
\vspace*{-0.4cm}
\caption[a]{Left: Centrality dependence of the charged particle multiplicity
  at midrapidity, $\eta=0$; Pb+Pb collisions at 2760~GeV. We compare
  our calculation for two UGDs to data by the ALICE collaboration.
  Right: Centrality dependence of the transverse energy at $\eta=0$.}
\label{fig:PbPb_Centrality}
\end{figure}
We shall focus on the centrality dependence of the charged particle
multiplicity at central rapidity, $\eta=0$, which we determine along
the lines described in section~\ref{sec:ktfact}. The result is shown
in Fig.~\ref{fig:PbPb_Centrality} for the UGDs with MV-model
($\gamma=1$) and AAMQS ($\gamma=1.101$ and $Q_s^2(x_0)\sim T_A$ or
$Q_s^2(x_0)\sim T_A^{1/\gamma}$) initial conditions.  We use $K=1.43$
for the former and $K=2.0$, 2.3 for the latter. The number of
final hadrons per gluon $\kappa_g=5$ in all three cases.

All UGDs give a rather similar centrality dependence of the
multiplicity which is in good agreement with ALICE
data~\cite{Aamodt:2010pb,Aamodt:2010cz}. On the other hand, they
differ somewhat in their prediction for the transverse energy. This is
of course due to the fact that the $\gamma=1.1$ initial condition
suppresses the high-$k_t$ tail of the UGD\footnote{One should keep in
  mind though that our estimate of the initial transverse energy
  carries a significant uncertainty of at least $\pm15\%$ related to
  our choice of $K$-factor; it is {\em not} determined accurately by
  the multiplicity since the latter involves only the product of
  $K$-factor and gluon $\to$ hadron multiplication factor
  $\kappa_g$.}. With the $K$-factors mentioned above these UGDs match
the measured $E_t$ in peripheral collisions. This is a sensible result
since one does not expect large final-state effects in very peripheral
collisions.  For the most central collisions the energy deposited
initially at central rapidity is about 0.5\% of the energy of the
beams and exceeds the preliminary measurements by
ALICE~\cite{Collaboration:2011rta} and CMS~\cite{Chatrchyan:2012mb} by
roughly 50\%.  This leaves room for $-p\, \Delta V$ work due to
longitudinal hydro
expansion~\cite{Gyulassy:1997ib,Eskola:1999fc,Dumitru:2000up}.

\section{Conclusions}

The upcoming p+Pb run at the LHC provides a new and unique opportunity
to study the dynamics of very strong color fields in nuclei at high
energies. The central question is whether QCD dynamically generates a
semi-hard scale $Q_s$ which dominates particle production and to test
its dependence on the valence charge density (resp.\ the nuclear
thickness) and on energy. In this paper we have provided
phenomenological predictions and expectations to the best of our
ability to model the large-$x$ valence degrees of freedom paired with
high-energy QCD evolution of the distribution of soft gluons in the
rcBK approximation. As we have discussed at length throughout the
paper, the CGC formalism at its present degree of accuracy is affected
by sizable uncertainties. Some are due to the lack of quantitative
control of higher order corrections to the formalism, while others are
related to the lack of data constraining the non-perturbative
parameters of the theory such as the initial conditions for the
evolution or the impact parameter dependence of the UGDs. The
forthcoming data will test whether the present formulation of the CGC
effective theory is quantitatively accurate or not.

\section*{Acknowledgements}
The research of J.~L. Albacete is supported by a fellowship from the
Th\'eorie LHC France initiative funded by the IN2P3. 
A.D.~is supported by the DOE Office of Nuclear Physics through
Grant No.\ DE-FG02-09ER41620 and by The City University of New York
through the PSC-CUNY Research Award Program, grant 65041-0043.\\
H.F.\ and Y.N.\ were supported in part by Grant-in-Aid for
Scientific Research (B) 22340064.\\
We
thank Paloma Quiroga-Arias for providing us with the EPS09 results
shown in Fig.~8 and also P.\ Tribedy and R.\ Venugopalan for providing
us with their CGC-predictions shown in Fig.~8 (top-left). Finally we
thank A.~Kovner and B.~W.~Xiao for very informative discussions on the
comparison of the inelastic and NLO corrections to the hybrid
formalism.



\end{document}